%% file: Paper.r1.tex
\documentclass[a4paper,twoside,10pt]{letter}
\usepackage{graphicx,saj,multicol,subeqnarray}
\usepackage{array}

\newcommand\citet[3][ et al.]{#2#1 (#3)}
\newcommand\citep[3][ et al]{(#2#1. #3)}

\DeclareRobustCommand{\ion}[2]{%
\relax\ifmmode
\ifx\testbx\f@series
{\mathbf{#1\,\mathsc{#2}}}\else
{\mathrm{#1\,\mathsc{#2}}}\fi
\else\textup{#1\,{\mdseries\textsc{#2}}}%
\fi}

\DeclareRobustCommand{\fion}[2]{[\ion{#1}{#2}]}

\newcommand\msun[1]{$#1\mathrm{M}_{\odot}$}

\newcommand\Ha{\mbox{H{$\alpha$}}}
\newcommand\Hb{\mbox{H{$\beta$}}}
\newcommand\Hi{\ion{H}{i}}
\newcommand\Hii{\ion{H}{ii}}
\newcommand\Hiir{\Hii\ region}
\newcommand\Oi{\fion{O}{i}}
\newcommand\Oii{\fion{O}{ii}}
\newcommand\Oiii{\fion{O}{iii}}
\newcommand\Nii{\fion{N}{ii}}
\newcommand\Sii{\fion{S}{ii}}

\newcommand\TG{NGC\,300}
\newcommand\wl[1]{$\lambda\;#1$~\AA}
\newcommand\wleq[1]{$\lambda = #1$~\AA}

\newcommand\dwleq[1]{$\Delta\lambda = #1$~\AA}
\newcommand\wll[2]{$\lambda\lambda\;#1, #2$~\AA}
\newcommand\linrat{\mbox{[\ion{S}{ii}]:\Ha}}
\newcommand\RA[3]{\mbox{$#1^\mathrm{h}#2^\mathrm{m}#3^\mathrm{s}$}}
\newcommand\RAd[4]{$#1^\mathrm{h}#2^\mathrm{m}#3^\mathrm{s}.#4$}
\newcommand\DEC[3]{\mbox{$#1^\circ#2^\prime#3^{\prime\prime}$}}
\newcommand\DECd[4]{$#1^\circ#2^\prime#3^{\prime\prime}.#4$}
\newcommand\Deg[1]{\mbox{$#1^\circ$}}

\newcommand\Sec[1]{\mbox{$#1^{\prime\prime}$}}

\newcommand\Units[1]{$\mathrm{#1}$}
\newcommand\deccolhead{${}^\circ{\;}^\prime{\;}^{\prime\prime}$}
\newcommand\fluxunits{\ifmmode\mathrm{erg\;cm^{-2}\;s^{-1}}\else$\mathrm{erg\;cm^{-2}\;s^{-1}}$\fi}

\newcommand\RatwErr[2]{\ifmmode{#1\pm#2}\else{$#1\pm#2$}\fi}

\newcommand\ATCA[2]{J00#1$-37$#2}
\newcommand\NS{N300-S}

\newcommand\Sim{$\sim$}

\def\papertype{Original Scientific Paper}

\def\udc{...}
\begin{document}


\markboth{\eightrm Extra-Galactic SNR Observing Techniques: NGC 300}
{\eightrm W. C. Millar, et. al.}

{\ }

\publ

\type

{\ }



\title{A Study of Optical Observing Techniques for Extra-Galactic Supernova Remnants: Case of \TG}


\authors{William C. Millar$^{1,2}$, Graeme L. White$^{3}$ and Miroslav D. Filipovi\'c$^{4,1}$}

\vskip3mm


\address{$^1$Centre for Astronomy, James Cook University, Townsville, Queensland 4811, Australia}
\address{$^2$Grand Rapids Community College, 143 Bostwick N.E., Grand Rapids, MI, 49503, USA}
\Email{wmillar}{grcc.edu}
\address{$^3$Charles Sturt University, Locked Bag 588, Wagga Wagga, NSW 2678, Australia}
\address{$^4$University of Western Sydney, Locked Bag 1797, Penrith South DC, NSW 1797, Australia}
\Email{m.filipovic}{uws.edu.au}


\dates{October 30, 2011}{February 13, 2012}

\summary{
We present the results of a study of observational and identification techniques used for surveys and spectroscopy of candidate supernova remnants (SNRs) in the Sculptor Group galaxy \TG. The goal of this study was to investigate the reliability of using $\linrat\geq0.4$ in optical SNR surveys and spectra as an identifying feature of extra-galactic SNRs (egSNRs) and also to investigate the effectiveness of the observing techniques (which are hampered by seeing conditions and telescope pointing errors) using this criterion in egSNR surveys and spectrographs. This study is based on original observations of these objects and archival data obtained from the Hubble Space Telescope which contained images of some of the candidate SNRs in \TG. We found that the reliability of spectral techniques may be questionable and very high-resolution images may be needed to confirm a valid identification of some egSNRs.
}


\begin{multicols}{2}
{

\section{1. INTRODUCTION}

The Earth is located in the gas and dust filled disk of the Milky Way galaxy (here after, the Galaxy) making the study of supernova remnants (SNRs) within the Galaxy difficult because the extinction and reddening effects of this interstellar medium (ISM) blocks or severely hampers our ability to see Galactic SNRs at wavelengths other than radio. For observations of SNRs in nearby galaxies this absorption by the ISM within both the host galaxy and the Galaxy is greatly reduced -- particularly for face-on (low inclination angle) spiral galaxies with high Galactic latitude (Matonick et al. 1997; Pannuti et al. 2000). Surveys of SNRs in the Local Group galaxies such as the Large and Small Magellanic Clouds (Filipovi{\'c} et al. 2005; Payne et al. 2008) and galaxies within some nearby clusters (mostly the Sculptor Group) have resulted in observations that are free from these biases. Over 450 SNRs have been found in nearby galaxies and listings are given by \citet{Matonick}{1997}, \citet{Uro{\v s}evi{\'c}}{2005}, \citet{Filipovi{\'c}}{2008} and \citet{Pannuti}{2007}.

An image of \TG\ is shown in Fig.~1. Table~1 provides a brief list of the characteristics of \TG\ following Millar et al (2011, here after, MWF11); a more complete list is provided by \citet{Kim}{2004}. \TG\ has a low inclination angle and a high Galactic latitude and observations of SNRs within \TG\ suffer very low internal extinction \citep{Butler}{2004} and foreground reddening\footnote{$E(B-V)=0.013$ mag \citep{Bland-Hawthorn}{2005}.}. \TG\ shows many giant \Hiir s which are evidence of many star formation episodes \citep{Read}{2001}. The similarities between \TG\ and other nearby spiral galaxies (such as M\,33 in the Local Group, and other members of the Sculptor Group, such as NGC\,7793) lead us to conclude that \TG\ is a typical, normal spiral galaxy (Blair et al. 1997, here after, BL97). Most authors have adopted distances of 2.0--2.1 Mpc for \TG\ (BL97; Payne et al. 2004, here after P04; Table~1) hence, we have adopted this distance of 2.1~Mpc to be consistent with previous SNR surveys and spectral observations (BL97; P04). The corresponding linear scale is 10.2~\Units{pc\;arcsec^{-1}}.

The resolving power (or seeing) of the two-meter class, ground-based telescopes usually used for SNR observations is generally from \Sec{0.5} (Chil\'e) up to \Sec{2} (Siding Spring, Australia or Sutherland, South Africa). These seeing conditions do not allow the direct imaging of extra-galactic SNRs (here after, egSNRs) and therefore, spectral line intensity ratios are used for identification of these objects. The now common technique of using two narrow band optical interference filters, one centered on \Ha\ (\wl{6564}) and the second allowing passage of the \Sii\ (\wll{6717}{6731}) doublet, to observe (or survey) bright nebulae in nearby galaxies was first described in a series of papers by \citet{Mathewson}{1972, 1973a,b,c}. This technique depends on the strength of the \Sii\ lines in SNRs being about the same strength as the \Ha\ line, which is most likely due to shock fronts in the expanding SNR shell as it collides with a dense ISM. In \Hiir s (where there are few if any shock fronts) this condition would not exist. The total flux of the two \Sii\ lines should be at least an order of magnitude weaker than the \Ha\ line in \Hiir s as compared to SNRs \citep{Mathewson}{1972}. Because of the filter's bandwidth, the \Ha\ filter was not able to remove the \Nii\ (\wll{6548}{6583}) doublet lines adjacent to the \Ha\ line. In some SNRs these lines (summed) can be as strong as the \Ha\ itself. Given this extra flux near the \Ha\ line, a candidate SNR was claimed if the emission region contained a (non-thermal) radio source and the \Ha\ + \Nii\ to \Sii\ ratio was less than two \citep{Mathewson}{1972}.

\citet{Dodorico}{1978} presents arguments based on observations of SNRs and \Hiir s within the Galaxy and with the Large Magellanic Cloud to show how SNRs can be identified within M\,33 when $\linrat > 0.4$. When $\linrat < 0.2$ the nebula is considered a \Hiir\ (BL97). When \linrat\ is between these two values the nature of the nebula may be unclear. \citet{Fesen}{1985} found that \Oi\ (\wll{6300}{6364}), \Oii\ (\wl{3727}) and \Oiii\ (\wll{4959}{5007}) are often all simultaneously strong in SNRs and this can also be used to help differentiate SNRs from \Hiir s in cases where \linrat\ is borderline.

The first observations of SNRs within \TG\ using this technique were published by \citet{Dodorico}{1980}. The candidates studied for this paper are the same candidates studied in MWF11 which were selected from those published by BL97 (optical candidates) and P04 (radio candidates). The positions of these SNRs and SNR candidates and the results of previous observations are given in Table~2 and are shown in Fig.~1. For Table~2, column~1 is the optical designation as given in BL97. Column~2 is the radio designation as given in P04. Columns~3 and~4 are the right ascension (RA) and declination (Dec) coordinates. Column~5 is the \linrat\ as reported in (or derived from) BL97. Column~6 is the \linrat\ as reported in MWF11 (including the measurement error discussed in MWF11). Column~7 is the spectral index ($\alpha$) as reported in P04 and column~8 is the measured diameter as reported in MWF11.

In MWF11, these candidates were studied with long-slit spectral observations and the accepted criteria ($\linrat\geq0.4$ and the presence of \Oi) were used giving a result of 22 objects as SNRs, with the remaining left as unknown (no signal) or unclear. For some of those objects which were unclear, the error in the \linrat\ measurement allows overlap with the $\linrat\geq 0.4$ threshold. The radio sources which do not reach the threshold value to be labeled as optical SNRs may be optically obscured by the emissions from neighboring \Hiir s (P04). This possibility is further supported by our investigation of the radio observations' astrometry here (section~3.1). There are other well established causes for the poor overlap of observed radio and optical SNR emissions \citep[]{Duric}{2000a,b}.

In this paper we investigate the reliability of using $\linrat \geq 0.4$ as a defining criterion for the detection and identification of egSNRs, and the effectiveness of the observing techniques used for egSNR surveys and spectrographs. We present a study of comparisons of the observations of these SNR candidates between BL97, P04, MWF11 and images containing the candidates found in the archives of the Hubble Space Telescope (HST). For reasons given below, this study included an analysis of the astrometry of the original CCD images from BL97, a discussion of the astrometry of the radio observations from P04, and an analysis of the pointing errors inherent in the two-meter class telescopes generally used for egSNR observations. Even with good seeing conditions, this telescope size class may be too small to ensure the reliability of using the \linrat\ ratio to identify egSNRs with complete confidence.

We found five sources which were comparable between the observation sets -- four radio sources and one optical source. We found an apparent systematic error in the radio observation astrometry. Unfortunately, with only four radio sources to work with in the current HST data, testing this possible error any further requires more high-resolution (HST) optical images. Only one source was in both the BL97 and HST observations. However this one source demonstrated a need for further investigation into the reliability of using the \Sii:\Ha\ line ratio for the identification of egSNRs in optical surveys and spectral observations.


\section{2. \TG\ SNRs IN THE HST ARCHIVAL DATA}

Table~3 shows the file names associated with the central wavelengths sorted by wavelength along with the original proposal identification and any publications based on the data. In this table, column~1 is the central wavelength of the HST filter, column~2 is the file name (all with \texttt{.FITS} extension) containing the image, column~3 is a list of publications associated with the observation. The references are included to provide the original objectives of the observations of \TG\ at these wavelengths and other publications which have used these observations (all taken from the MAST\footnote{\texttt{http://archive.stsci.edu/}} web site).

Table~4 shows the bandwidths for each of the HST filters used for these images. These data were taken from the images' FITS file header\footnote{They are also available on the HST WFPC2 instrument website. \texttt{http://www.stsci.edu/hst/wfpc2/documents/IHB\_17.html}}. Column~1 is the filter's central wavelength and column~2 is the filter bandwidth. This table also shows emission wavelengths with the associated ions or atoms that are found within the bandwidth of the filters (column~3). These ions or atoms are known to be within (Galactic) SNRs \citep{Fesen}{1996}.

A computer program was written to read the HST FITS file headers and based on each file's FITS world coordinate system (WCS) the following objects (from Table~2) were found in one or more of the images files: \ATCA{5438}{4144}, \ATCA{5440}{4049}, \ATCA{5445}{3847}, \ATCA{5450}{4030}, \ATCA{5450}{3822}, \ATCA{5450}{4022}, \ATCA{5451}{3826}, \ATCA{5451}{3939}, \ATCA{5500}{4037}, \ATCA{5501}{3829}, \ATCA{5503}{4246}, \ATCA{5503}{4320}, \NS8, \NS9, \NS10, \NS13, \NS14, \NS15, \NS16, \NS18.

Table~5 lists the filter central wavelengths and bandwidths containing spectral emissions which are diagnostic to the identification of SNRs (see discussion in section~1). This table's columns are the same as those in Table~4. The emissions of most interest are: \Ha\ (\wl{6565}), \Sii\ (\wl{6717}, \wl{6732}), \Oi\ (\wl{6300}), and \Oiii\ (\wl{5007}). Only the files possibly containing these wavelengths were examined any further. As can be seen from Table~3, only one image file group contains data with \Ha\ (\texttt{u671370\#r\_drz}, where \texttt{\#} is one of the digits, 5 to 9). There were no files containing the \Sii\ doublet wavelengths. The images centered on \wleq{5741} and \wleq{6001} contain wavelengths of interest but none of these images contain objects from Table~2. Thus, only the \Ha\ images were of use for this study. Only 5 of the 51 objects were found within the \Ha\ images. All of the \Ha\ images have the same field of view. The final results of the search are shown in Table~6. The columns fro this table are the same as those for Table~2.

The HST \Ha\ images are centered on (\RAd{00}{54}{54}{54}, \DECd{-37}{40}{35}{9}). They are 400~second exposures taken in May of 2001 with the Wide-Field Planetary Camera 2 (WFPC2) instrument using its (default) PC1 aperture. The filter used in these images (F656N) was centered on \wleq{6563.76} with a RMS bandwidth of \dwleq{53.77}. The bandpass characteristics of this filter are shown in Fig.~2. The data were originally collected as part of the HST proposal 8591 by, Douglas Richstone \citep[]{Richstone}{2000}. The file \texttt{u6713709r\_drz.fits} appeared to provide the cleanest image for analysis of the SNR candidates' environment.

Figure~3 shows the image contained in this file in negative gray-scale. The five SNR candidates (Table~6) in the field of view are labeled and positioned by DS9 pandas. For comparison, Fig.~4 shows an image with the same position, size and rotation using DSS2-Red data from SkyView\footnote{\texttt{http://skyview.gsfc.nasa.gov/}}. There is a slight difference in plate scale between the two images. The green line in Fig.~4 is the outline of the HST image's (Fig.~3) first quadrant, which contains the high-resolution CCD chip and the center of \TG. A bright star in the second quadrant which appears in the DSS2-Red image does not appear in the HST image (the DSS2-Red survey includes \Ha\ wavelengths). There are some differences in the apparent distribution of gas in the DSS2-Red image compared to the HST image, probably due to the wider pass band of the DSS2-Red image. There is a triangular shaped optical artifact located to the north of the very bright star near the bottom of the DSS2-Red image. The four radio SNR candidates from P04 were not optically confirmed, the optical SNR candidate from BL97 was confirmed (MWF11, Table~2).


\section{3. ANALYSIS OF THE RADIO SOURCES}

When the HST image was displayed in false color\footnote{Using the Smithsonian Astrophysical Observatory's DS9 software package \texttt{http://hea-www.harvard.edu/RD/ds9/}.} to show emission intensity (Figures~5 through 8), a number of small sized, high intensity objects appeared in the image. Most likely these are stars but that is difficult to determine from the available data. Here after these objects are referred to simply as ``hot spots.''

Multi-frequency observations of egSNRs are limited by current technologies to radio, optical and X-ray emissions. None of the four radio sources found in this HST image are known to have emissions in other wavelengths. NS16 is observed only in optical.

\subsection{3.1 Radio Observations Astrometry}

The majority of the SNRs found in the Galaxy and the Magellanic Clouds are 50~pc or less in diameter \citep[]{Strom}{1996}. If the same is assumed for \TG, then the SNRs in \TG\ should be less than \Sec{5} in diameter (at the assumed distance). The radio observations published in P04 used a circular beam width of \Sec{6}. Thus the typical SNR would not fill the beam of the radio telescope. This may have led to systematic astrometric errors.

In Figures~3 through 8 the images (particularly the HST \Ha\ images) appear to show that there is such a systematic error in the astrometry of the radio candidates. The candidates show a consistent displacement to the southwest from the neighboring \Hiir. To account for radio emissions typical of a SNR without corresponding optical emissions (particularly when the candidate is $\sim$50~pc in diameter) the radio source should be located within or adjacent to the H cloud region so the candidate SNR's optical emissions could be masked. Though this is not the only possible explanation for such candidate SNR radio/optical observations (Duric. 2000a,b; Pannuti et al. 2000), it is likely (P04). No ground-based telescope of less than four meters should be able to separate this candidate from the \Hiir.

Radio observations with a higher resolving power or very high-resolution optical observations (HST) of the other radio sources are needed to resolve this possible apparent error.

\subsection{3.2 SNR Candidate \ATCA{5450}{4030}}

This radio SNR candidate has a spectral index of \RatwErr{-0.5}{0.2}. It has a \linrat\ of \RatwErr{0.32}{0.12} (MWF11, Table~2). Because the threshold value of the \linrat\ is 0.4 this candidate could not be classified as an optical SNR (MWF11). Note however that the threshold value is within the error of the line ratio measurement. Figure~5 shows the detail in the HST image at the location of this radio source. The candidate is located very near the edge of a \Hiir\ (\Hii\ C29, Soffner et al. 1996, \Sec{8.33}, 85~pc away from center) showing no special structural properties.

A single image in \Ha\ does not allow us to determine the nature of the emissions from the region. An HST \Sii\ image would be very helpful. If there is a systematic error in the astrometry of the radio source, it is possible that the radio source is actually inside the \Hiir\ and the \Hiir\ would then interfere with the optical line ratio measurement. There are many known SNRs without optical emission so some of the \TG\ SNRs would be expected to be in that group. 

\subsection{3.3 SNR Candidate \ATCA{5450}{4022}}

Figure~6 shows detail in the HST \Ha\ emissions surrounding the radio source \ATCA{5450}{4022} which was classified as being either a SNR or a \Hiir\ in P04. The radio source is located on the edge of a large bubble in the \Hiir\ (76C, Deharveng et al. 1988, \Sec{2.36}, 24~pc away from center). The small magenta (\Sec{1}) circle pointed out by the red arrow shows the position of the candidate. The larger magenta ellipse (at the edge of which the radio candidate is located) outlines one possible border to the large bubble. A smaller region within the ellipse which can be seen at the green emission level may also be a SNR shock front within this \Hiir. Without proper \fion{S}{ii} data these remain simply possibilities of shock fronts.

The magenta ellipse is \Sec{2.54}$\times$\Sec{2.38} or $25.9\times24.8$~pc with its semi-major axis aligned in declination. This is approximately the size of an SNR of about 10\,000 to 20\,000 years in age \citep[]{Cioffi}{1990}. SNRs of this age are typically in the radiative, pressure-driven snowplow (PDS) stage which likely presents a bubble shaped appearance such as the one seen in this image. The \linrat\ for this candidate is \RatwErr{0.38}{0.31} (MWF11, Table~2). While the ratio is just below the critical value of 0.40, the large noise error in this measurement combined with the apparent structure next to the radio candidate raises serious concern about denying its classification as an optical SNR. This evidence is only available because of this high resolution image from the HST.

Also shown in Fig.~6 is a black line corresponding to the diameter measurement of this candidate (130~pc) as published in MWF11. This allows comparison of the poorer seeing conditions during those observations with the resolution of the HST image. This 130~pc (\Sim{\Sec{12.7}}) line is drawn centered on the published position of the radio source. Given this diameter comparison, the \linrat\ may be that of the \Hiir\ with a large error caused by seeing conditions. No firm conclusion about the optical emissions of this radio source can be drawn from these data with the seeing conditions degrading the line ratio technique by increasing the measurement error.

\subsection{3.4 SNR Candidate \ATCA{5451}{3939}}

Figure~7 shows the location of \ATCA{5451}{3939}, next to an \Hiir\ (\#20, Bresolin et al. 2009, \Sec{3.80}, 39~pc away from center). This \Hiir\ has a bright emission source off to one side at the same location as the radio source. It is not likely this ``hot spot'' could be the SNR but it could be the known Wolf-Reyet (WR) star from (\#18, Schild et al. 2003). There is also a known star association in this region (AS 57, Pietrzy{\'n}ski et al. 2001). Emissions from these stars (the association or the WR) can not be separated from the SNR candidate in the spectrographic slit and the WR star may also contribute to the radio emissions (we note the discussion of excessive radio emissions and the spectral index of Galactic WR stars in Montes et al. (2009)).

This radio source could be a SNR within the \Hiir\ or it could be a coincidence with a combination of stars. Observations with the HST's WFPC2 instrument using the F673N filter to indicate the presence of any \Sii\ emission could be very helpful here. The spectroscopic measurements from MWF11 do not support this object as an SNR.

\subsection{3.5 SNR Candidate \ATCA{5500}{4037}}

Figure~8 shows the location of the SNR candidate \ATCA{5500}{4037}. The radio source is outside of any notable \Ha\ emission. There is a large inset into the \Hiir\ (\Hii\ E18, Soffner et al. 1996, \Sec{4.45}, 45~pc away) on the side nearest the radio source. However, if this inset were due to a supernova event the SNR would definitely be in the PDS stage and both radio and optical emissions should be coming from the shock front, not from the center of, or outside of the expanding shock. There are at least four known \Hiir s, a planetary nebula and a star association in the vicinity of this radio source but they are far enough from the source that they should be separable in the spectrometer slit -- given the correct source astrometry and telescope pointing. There is a possibility that this radio source is a background object (e.g. a quasar) which happens to lie very near a \Hiir\ within \TG\ although this source was classified as an SNR as opposed to SNR/\Hiir\ in P04. As has been mentioned previously, this could also be a SNR without optical emissions.

If the apparent systemic radio astrometry error is real, it may be an SNR with its optical emissions masked by the \Hiir\ emissions. The emissions from the SNR and the \Hiir\ would be impossible to distinguish with a two-meter class telescope.


\section{4. ANALYSIS OF THE OPTICAL CANDIDATE}

The \NS16 \linrat\ ratio was not measured from spectra in BL97 but Table~3A of that paper shows \NS16\ with a \linrat\ of 0.70, based on the BL97 survey CCD images. BL97 also reports a (\Ha) diameter of 52~pc. From MWF11, \NS16 has an observed \linrat\ of \RatwErr{0.94}{0.06} and a measured diameter of 57~pc. This large measured ratio is based on fairly low flux levels: \Ha\ = 2.2 and \Sii\ = 2.1($10^{-15}\;\fluxunits$).\footnote{The BL97 candidates' spectral flux density was consistently weaker by an order of magnitude compared to the P04 candidates. (See MWF11.)} \citet{Hoopes}{1996} demonstrated that a \linrat\ of up to 0.5 can be created by the diffuse ionized gas (DIG) of \TG. Because this candidate is not near any known \Hiir\ this ``background ratio'' could be subtracted from the MWF11 ratio and the result still exceeds the 0.4 critical \linrat\ ratio for indication of an SNR. However, the critical value is then within the small noise error.

Figure~9 shows the BL97 images (set G) of \NS16 zoomed and cropped. On the left is the \Ha\ filter image and on the right is the \Sii\ filter image. In both images only a faint increase from the background is seen at \NS16's given position. The inner panda circle is centered on the \NS16 coordinates with an apparent \Sec{8} (81.6~pc) diameter -- somewhat larger than the measured diameter. The outer panda circle is twice as big.

Figure~10 zooms in to the HST \Ha\ image at \NS16. In this figure there is no evidence of any structure in the \Ha\ emissions. It is possible that the \linrat\ test for SNR character has failed. No satisfactory explanation of the apparent contradiction between the two observations has been found.


\section{5. THE ASTROMETRY OF BL97}

There is contradictory evidence for the existence of a SNR at the location labeled as \NS16. BL97 and MWF11 have high \linrat\ values but the HST \Ha\ image shows no indication of any structure in the vicinity. In an attempt to find a resolution we checked the astrometry of the BL97 CCD images. \label{mark.bl97astro}The CDD images used for the BL97 optical survey were provided by, William Blair. These images were sets and each set was labeled with the letters `C' through 'J.' Each lettered image set contained a continuum image, a \Ha\ filter image and a \Sii\ filter image with the same center, plate scale and rotation (see BL97). The files did not include the `I' image (which was used in BL97) but did include a `K' image which was not used in BL97. The `I' image contained SNR candidates \NS12 and \NS17. The `C' and `K' images contained no SNR candidates and thus were not considered in this analysis.

The images are in the flexible image transport system (FITS) format, 800 pixels on a side, with a field of view of approximately \Deg{0.087}. The central coordinates of each image are given in Table~2 of BL97. The image files did not have a FITS WCS and the $x$-axis needed to be inverted to obtain the correct orientation (north up and east to the left). A WCS was thus created for each of the FITS files.

\subsection{5.1 Finding Known Objects Within the Images}

A list of 2MASS point sources from the online catalog was created from of a search for all sources within a radius of one arc-degree of the center of \TG. The program used a dynamic array of a record structure to store the coordinates of each 2MASS point source along with a ``distance'' data field, which is discussed below. NASA's SkyView was used to create DSS and DSS2-Red FITS file images which appeared equivalent to the BL97 images. The SkyView 2MASS images were used as a visual aid in associating bright, easy to locate 2MASS point sources with optical counterparts in each BL97 image. To reduce errors in calculating the image scale the chosen sources were always closer to (but not at) the edge of the image.

The BL97 ($x,y$) coordinates and 2MASS ($\alpha, \delta$) coordinates of each of the three 2MASS sources from each image were entered into a computer program. The estimated coordinates were compared to the list of catalog coordinates. The comparison created a distance measurement between the estimated and the catalog coordinates which was stored in the ``distance'' data field of each point source data record structure. The source list was then sorted on the distance field and the source with the smallest distance was put at the top of the sorted list. The catalog coordinates at the top of the list were the coordinates actually used for calculating the BL97 image center, plate scale and rotation.

\subsection{5.2 Program Results and Astrometry of the SNRs}

Table~7 shows the program results for the calculation of the image centers. Column~1 is the image identification letter. Column~2 is the image center as given in BL97's Table~2. Column~3 is the calculated center from the program and column~4 is the distance (or offset) between them, in arcseconds.

After studying the continuum, \Ha\ filter and \Sii\ filter images for each lettered image set, any difference in the center coordinates between the image filter types appeared to be negligible. Because the filters used to make the \Ha\ and \Sii\ images where narrow band, it was impossible to locate 2MASS point sources within those images. Thus, center coordinates and plate scale values were calculated only for the continuous images and then applied to all image filter types for the image field (lettered set). Based on the calculated WCS for each image, Table~8 shows the reported equatorial and calculated pixel coordinates for each of the BL97 SNR candidates.

\NS16 is the only BL97 SNR candidate to appear in an HST image. Figure~11 shows the BL97 \Ha\ filter and \Sii\ filter images of \NS1 (image set~D). The pandas mark the coordinates given for \NS1 in BL97 as calibrated by 2MASS point sources. The BL97 astrometry is good -- certainly within the seeing conditions reported in BL97 (\Sim{\Sec{1}}). Figure~12 shows a comparison of the reported position (BL97, on left) of \NS16 with the apparent center of the assumed image of the candidate (HST, on right). The offset in these positions (\Sec{0.7}) is again within the reported seeing conditions. The logical next step would be to calibrate the HST images to 2MASS and study the HST \Ha\ image of \NS16\ in comparison to the BL97 images.


\section{6. HST IMAGE ASTROMETRY}

The HST FITS file headers indicated that the HST Guide Star Catalog (GSC) version 1.x was used for the image astrometry. According to the Astrometry Department of the United States Naval Observatory\footnote{\texttt{http://ad.usno.navy.mil/star/star\_cats\_rec.shtml}} the accuracy of this catalog is approximately 500~mas. The biggest problem with the GSC is that it does not provide for the proper motion of the stars.

The attempt to calibrate the HST images to the 2MASS catalog using the procedure and software from the BL97 analysis proved to be impractical because of the high-resolution of the HST images. The difference in resolution between the SkyView 2MASS images and HST images made it impossible to correctly identify 2MASS point sources in the HST images with reasonable confidence. Therefore we used an alternative method to check the astrometry.

\subsection{6.1 Alternate Analysis}

An analysis of the HST and DSS2-Red files' WCS accuracy against the BL97 2MASS-calibrated images was made by comparing the position of a bright star which could be identified in all image sets. The first comparison, between BL97 and the HST images, is shown in Fig.~13. On the left is the BL97 G~\Ha\ image containing \NS16 and on the right is the HST \Ha\ image containing \NS16. The position of the selected star within these images is listed in Table~9. In this table column~1 is the image used for the observation (BL97, HST, DSS2-Red), columns~2 and 4 are the J2000 RA and Dec coordinates of the center of the image of the star as found using DS9. Table~9 columns~3 and 5 are the difference between the positions (with BL97 as the reference) in arcseconds. Column~6 is then the difference in position (offset) between the two images, measured in arcseconds. Between the BL97 and HST images there is an offset of \Sec{0.15}.

The second comparison was made between BL97 and the DSS2-Red images used for the finding maps for the observations in MWF11 and is shown in Fig.~14. Between the BL97 and DSS2-Red images there was an offset of \Sec{0.13}. \citet[]{Mickaelian}{2004} measured the astrometry of the DSS2-Red images based on AGN positions and found an accuracy of \Sec{0.33}. These comparisons demonstrate that any error in the astrometry of the images used in BL97 was within the reported seeing conditions.

\subsection{6.2 Estimating Positional Error}

Because of the difficulty in recalibrating the HST image WCS the alternative procedure of convolving the position errors was used. In section~5.2 it was found that the position error in the BL97 images was $\sim$\Sec{1} (mainly due to seeing). This was convolved with the approximate positional error of the HST images from Table~9 to find a positioning error of the BL97 SNRs on the HST images: The result is \Sec{1.0} -- essentially the seeing conditions of BL97. The same was done with the radio SNRs from P04. According to Table~2 of P04 these candidates were observed with a circular beam width of \Sec{6}. Convolving 10\% of that gives a $1\sigma$ positioning error for the candidates as \Sec{0.62} or \Sec{0.6}.

These values were used as the $1\sigma$ positioning error for finding the SNR candidates within the HST images. We see that positioning (astrometry) errors are due mainly to seeing conditions and telescope pointing.


\section{7. SEEING AND TELESCOPE POINTING}

Some of the candidates from BL97 which were observed in MWF11 returned no signal. In an attempt to determine the cause, the telescope pointing accuracy was analyzed by superimposing the slit camera images on top of the observation finding maps. Figures~15 and 16 show examples of the results. Figure~15 shows a very good alignment with \NS2. In Fig.~16 there is a confusion of sources surrounding \NS11. Both \NS11 and \ATCA{5442}{4313} were clearly within the slit and the seeing conditions were undoubtedly allowing \ATCA{5443}{4311} to also flow through. Exactly what was being measured here is uncertain but it was labeled as \NS11 (MWF11) because it was the principle candidate for that observation. This observation also included emission from the neighboring \Hiir.

For \NS11, BL97 reports $\linrat=0.66$ (based on CCD images, not spectra) and MWF11 reports \linrat\ = \RatwErr{0.30}{0.12}. Because of source confusion, neither of these measurements can be trusted. BL97 claimed seeing of \Sec{1}, but these sources would be confused in any spectrometer's slit. Due to seeing conditions creating telescope positioning error and the astrometry error in the radio observations it is not possible to determine if \ATCA{5442}{4313} and \NS11 are actually the same object. Measurements of the diameter of \NS11 (BL97, MWF11) are greater than 100~pc. This large diameter may be caused by source confusion with the \Hiir\ or it may be a multiple supernovae site. Unfortunately multiple SNR sites a usually linked to OB associations and there are no known OB associations in this region. The only way to resolve this problem is with a higher-resolution telescope with no seeing problems -- the HST. Unfortunately no archival HST file contained an image of these candidates.

Figure~17 shows the slit camera image on the finding map of \NS16 used for the observations for MWF11. For this candidate the telescope pointing was erroneous (by \Sec{1} to \Sec{2} or 10 to 20~pc) and yet a \linrat\ \RatwErr{0.94}{0.06} was measured. The cause of the poor alignment is difficult to determine -- equipment or seeing conditions. It is most likely due to seeing conditions which were on the order of two to three arcseconds for most nights of the observing run. The seeing conditions may have allowed most of the flux from \NS16 through the slit but it is then difficult to account for the higher ratio compared to the BL97 results (0.70).

However, we still have the real problem with this observation -- the HST image containing \NS16 shows no evidence of the candidate's existence (Fig.~10).


\section{8. ANALYSIS OF THE BL97 IMAGES}

\label{mark.ns16}BL97 obtained long slit spectra for \Sim{$\frac23$} of the candidates found in the optical filter survey for that paper. With 100\% confirmation ($\linrat>0.4$) of this $\frac23$ sample, there was confidence in all 28 candidates being confirmed as SNRs \citep[]{Long}{1996}. \NS16 was not a member of that $\frac23$ subset.

\subsection{8.1 Candidate Image Profiles}

Optical emissions are expected to be greatest during the PDS stage \citep[]{Cioffi}{1990}. This stage is typically 50~pc at maximum diameter. At the distance to \TG\ 50~pc is equivalent to \Sec{4.9}. The images for BL97 are $800\times800$ pixels at 5.3~arcminutes square (BL97; Long. 1996) and the pixel size is then \Sec{0.4}. At the distance to \TG\ this is about 4~pc. The maximum theoretical size of a PDS stage SNR on a BL97 image would be about 12 pixels in diameter or an area of about 110 pixels squared.

The \Ha\ and \Sii\ images used for BL97 were analyzed for candidate image profiles with the intention of investigating the actual number of CCD pixels used to determine the \linrat\ ratio of the candidate SNRs. Using the ``Co-Add'' command of SBIG's \emph{CCDOps} program\footnote{Available from Santa Barbara Instruments Group, \texttt{http://www.sbig.com/sbwhtmls/ccdopsv5.html}} the image of each SNR candidate was stacked to form a composite image. This was done with both the \Ha\ and \Sii\ images. The $x$ and $y$-axis profiles of these composite images are shown in Figures~18 and 19.

Figure~18 shows two renditions of the stacked \Ha\ candidate images. These two renditions differ by the display intensity scale and the number of plotted contour lines. The images were stacked so all SNR candidates were located (the pixel corresponding to their equatorial coordinates) at the same resultant pixel. The rendition on the right has the crosshairs and profiles centered on the emissions peak which happens to be the same as the coordinate pixel. As an example of the data, Table~10 shows a $9\times9$ grid of the CCD pixel values from Fig.~18. Figure~20 shows a surface plot of these values which clearly shows a high signal level against the background. Figure~21 shows a line plot the values along the $x$-axis (E-W) at $y$-axis (N-S) row number 165. The full-width half maximum is about $10\times12$ pixels (\Sim{$41\times49$}~pc) which is about the same as the expected 12 pixel diameter area and the most expected diameter for candidates at this stage of their evolution.

In Fig.~19 the \Sii\ frames were stacked but did not include the H and J images as these were too noisy for this procedure. Various techniques were used to remove the noise\footnote{\emph{CCDOps} has some noise treatment routines such as, ``Kill Hot Pixels,'' ``Kill Cold Pixels,'' ``Smooth Pixels.''}. The problem was that the \Sii\ signal on these images was too close to the noise floor. The H and J images contained candidates: \NS2, S3, S4, S7 (H image); S20, S26, S27 (J image). Figure~19 shows two renditions of the stacked \Sii\ images. On the left the crosshairs (and the profiles) are centered on the coordinate pixel (508, 164). On the right, the crosshairs (and profiles) are centered on the pixel with maximum value (509, 162). This difference in pixel position corresponds to about 4 (RA) and 8~pc (Dec) in physical distance, respectively. The SNR positions were determined from the \Ha\ images based on \Hiir s but this also implies that the maximum \Ha\ and the maximum \Sii\ emissions are on the opposite side of the SNR. The surface plot of the \fion{S}{ii} emissions shown in Fig.~22 shows multiple peaks distributed around the centers of the SNR candidates. The FWHM (centered on the greatest peak) is about $9\times9$ pixels (\Sim{$37\times37$}~pc).

The estimated FWHM of \Ha\ and \Sii\ profiles were convolved with the FWHM of a small star from the \Sii\ J image. The profile of the star is shown in Fig.~24. When the FWHM of the stacked \Sii\ image profile (Fig.~23) was convolved with the small star image profile (Fig.~24) the resulting apparent size of the stacked \Sii\ regions was,
\begin{equation}
\sqrt{(28\;\mathrm{pc})^2 - (15\;\mathrm{pc})^2} = 23\;\mathrm{pc} = 6\;\mathrm{pixels}
\end{equation}
This was taken as the average SNR candidate diameter, so the extant image area of the \Sii\ emissions was approximately 28 pixels squared. Because the images could not be flux calibrated, the number of pixels actually contributing to the \linrat\ could not be estimated with reasonable confidence. The theoretically expected maximum number of pixel was about 110. With only 28 pixels (in a stacked image of all sources) contributing to the flux measurement of spectra, flux measurements of BL97 may well have large errors. Weak signals may lead to large flux measurement error (as an overestimate) an effect noted in other egSNR observations by \citet{Blair}{1981}.


\subsection{8.2 Images of \NS16}

Figure~11 shows the BL97 \Sii\ and \Ha\ images zoomed in on \NS16. As measured in MWF11 the \Ha\ and \Sii\ flux levels for \NS16 were about $2\times10^{-15}$~\fluxunits\ which was only about 100 CCD pixel counts above the background. The spectrum shows a high-level background along the slit. Figure~17 shows the slit was at least an arcsecond away from the reported coordinates of the candidate. The spectrum from MWF11 of \NS16 is shown in Fig.~26.

In the HST \Ha\ image of \NS16 (Fig.~10) there was little if any evidence of a disturbance in the H gas in the region of the location of \NS16. While the equivalent images of the radio SNR candidates (see Figures~5, 6, 7 and 8) show strong evidence of such, even though the measured \linrat\ ratio does not confirm them as optical SNRs. There is also no known X-ray emission associated with \NS16.

Figure~27 shows a 3-D plot of the BL97 CCD \Ha\ (top) and \Sii\ (middle) image pixel counts of \NS16. The pixels plotted (a 27 $\times$ 27 array centered on the candidate's coordinates) are shown by the green box in the image tile on the left. The plot is shown on the right. In both cases the emission is only slightly above the background but clearly discernible. The size of the candidate (50~pc) implies it to be at the end of its PDS stage. The 3-D plot on the bottom of the figure is a plot of the result of dividing the \Sii\ pixel value by the \Ha\ pixel value. The apparent SNR disappears in the high-level DIG noise.

There are spectral line characteristics other than \linrat\ to support SNR candidacy (e.g. the presence of \Oi\ in the spectrum) and there is a selection effect biased toward finding SNRs away from H regions in optical surveys (P04). Type~Ia SN which are away from any H region are then located in less dense ISM. As a result, there is a lower abundance of shocked material to produce the \fion{S}{ii} (and other metallic) spectral lines in the SNR remnant. Such SNR spectra may be dominated by the Balmer lines (e.g. SN1006, Tycho's and Kelper's SN) and are generally missed by optical surveys \citep{Pannuti}{2000}. A check of the data from MWF11 shows no Balmer dominance in the spectrum of \NS16, particularly with a \linrat\ of 0.6 and a high level of \fion{O}{i} usually associated with shocked ISM. High-resolution imaging should be used to confirm the existence of this SNR. This imaging could be done with space-based telescopes (HST) or with four-meter class telescopes. Better results may be obtained with ground-based telescopes using adaptive optics.

\section{9. CONCLUSION}

A careful investigation of the data collected with the telescopes and instruments typically used for the discovery and confirmation of extra galactic supernova remnants reveals that the reliability of these techniques may be questionable. A strict flux density measurement error analysis shows that large errors in the \linrat\ ratio occur when the ratio is based on low flux density levels (MWF11). Seeing conditions lead to blurring of telescope positioning and thus pointing errors which may impinge in the reliability of the flux measurements and on the confidence of exactly what object is being measured (Fig~17).

The seeing conditions also introduce error into the astrometry of the sources which is generally as large as the telescope pointing error. If these errors add in the same direction there is a possibility that the spectrograph slit is completely off the candidate. In cases where the target galaxy has high level emission from its diffuse ionized gas (such as \TG), the spectroscopic signal may not be sufficient to confirm the presence of a supernova remnant. High resolution optical images from space-based telescopes or from ground-based telescopes using adaptive optics may be necessary for confirming the existence of these extra-galactic supernova remnants.


\acknowledgements{
We gratefully acknowledge the generosity of Professor William Blair who provided us with the original CCD image files from the BL97 survey observations. We also thank the Hubble Space Telescope archival team for their work in maintaining the archives thus making the telescope's original data available to all.}

\include{PaperBib}

}
\end{multicols}

\include{Figures}

\include{Tables}


\vfill\eject

{\ }



\naslov{OPTIQKE TEHNIKE POSMATRA{NJ}A VAN GALAKTIQKIH OSTATAKA SUPERNOVIH: SLUQAJ \TG}


\authors{William C. Millar$^{\bf 1,2}$, Graeme L. White$^{\bf 3}$ and Miroslav D. Filipovi\'c$^{\bf 4,1}$}

\vskip3mm


\address{$^1$Centre for Astronomy, James Cook University, Townsville, Queensland 4811, Australia}
\address{$^2$Grand Rapids Community College, 143 Bostwick N.E., Grand Rapids, MI, 49503, USA}
\Email{wmillar}{grcc.edu}
\address{$^3$Charles Sturt University, Locked Bag 588, Wagga Wagga, NSW 2678, Australia}
\address{$^4$University of Western Sydney, Locked Bag 1797, Penrith South DC, NSW 1797, Australia}
\Email{m.filipovic}{uws.edu.au}

\vskip3mm


\centerline{\rrm UDK \udc}

\vskip1mm

\centerline{\rit Originalni nauqni rad}

\vskip.7cm

\begin{multicols}{2}

{


\rrm

U ovoj studiji predstav{lj}amo posmatraqke tehnike korix{cc}ene za spektroskopsku identifikaciju kandidata ostataka supernovih u \TG\ iz Skulptor Grupe galaksija. Glavni ci{lj} ove studije odnosi se na ispitiva{nj}e svrsishodnosti korix{cc}e{nje} parametra $\linrat\geq0.4$ u identifikaciji van-galaktiqkih ostataka supernovih. Posebna pa{zz} nja posve{cc}ena je qi{nj}enici da posmatraqki uslovi kao sto su vidljivost i poziciona preciznost teleskopa imaju dominantnu ulogu u korix{cc}e{nj}u ove tehnike. Ova studija bazira se na arhivskim posmatra{nj}ima sa {\rm  Hubble Space Telescope} na kojima su identifikovani ostatci supernovih u Skulptor galaksiji --- \TG. Naxi rezultati pokazuju da je primen{lj}ivost ove tehnike veioma nepouzdana i da posmatra{nj}a u vixoj rezulucijic su neophodna prilikom identifikacije van-galaktiqkih ostataka supernovih.

}

\end{multicols}

\end{document}

%% file: PaperBib.tex

\newcommand\rmxaa{Rev. Mexicana Astron. Astrofis.}
\newcommand\asph{Astrophysics}
\newcommand\al{Astronomy Letters}
\newcommand\aj{Astron.~J.}
\newcommand\apj{Astrophys.~J.}
\newcommand\apjl{Astrophys.~J. Lett.}
\newcommand\apjs{Astrophys.~J. Suppl. Ser.}
\newcommand\apss{Astrophys. Space Sci.}
\newcommand\aap{Astron. Astrophys.}
\newcommand\aaps{Astron. Astrophys. Suppl. Ser.}
\newcommand\an{Astronomische Nachrichten}
\newcommand\cjaap{Chinese~J. Astron. Astrophy.}
\newcommand\mnras{Mon. Not. R. Astron. Soc.}
\newcommand\pasp{Publ. Astron. Soc. Pac.}
\newcommand\pasa{Proc. Astron. Soc. Aust.}
\newcommand\nar{New Astron. Rev.}
\newcommand\sci{Science}

\references

Andersen, D.R., Walcher, C.J., B{\"o}ker, T., Ho, L.C., van der Marel, R.P., Rix, H., Shields, J.C.: 2008, \journal{\apj}, \vol{688}, 990.

Barth, A.J., Strigari, L.E., Bentz, M.C., Greene, J.E., Ho, L.C.: 2009, \journal{\apj}, \vol{690}, 1031.

Beifiori, A., Sarzi, M., Corsini, E.M., Dalla Bont{\`a}, E., Pizzella, A., Coccato, L., Bertola, F.: 2009, \journal{\apj}, \vol{692}, 856.

Berger, E., Soderberg, A.M., Chevalier, R.A., Fransson, C., Foley, R.J., Leonard, D.C., Debes, J.H., Diamond-Stanic, A.M., Dupree, A.K.,  Ivans, I.I., Simmerer, J., Thompson, I.B., Tremonti, C.A.: 2009, \journal{\apj}, \vol{699}, 1850.

Blair, W.P., Long, K.S.: 1997, \journal{\apjs}, \vol{108}, 261.

Blair, W.P., Kirshner, R.P., Chevalier, R.A.: 1981, \journal{\apj}, \vol{247}, 879.

Bland-Hawthorn, J., Vlaji{\'c}, M., Freeman, K.C., Draine, B.T.: 2005, \journal{\apj}, \vol{629}, 239.

B{\"o}ker, T., Lisenfeld, U., Schinnerer, E.: 2003a, \journal{\aap}, \vol{406}, 87.

B{\"o}ker, T., Stanek, R., van der Marel, R.P.: 2003b, \journal{\aj}, \vol{125}, 1073.

B{\"o}ker, T., Laine, S., van der Marel, R.P., Sarzi, M., Rix, H., Ho, L.C., Shields, J.C.: 2002, \journal{\aj}, \vol{123}, 1389.

B{\"o}ker, T., Sarzi, M., McLaughlin, D.E., van der Marel, R.P., Rix, H., Ho, L.C., Shields, J.C.: 2004, \journal{\aj}, \vol{127}, 105.

Bond, H.E., Bedin, L.R., Bonanos, A.Z., Humphreys, R.M., Monard, L.A.G.B., Prieto, J.L., Walter, F.M.: 2009, \journal{\apjl}, \vol{695}, 154.

Bresolin, F., Pietrzy{\'n}ski, G., Gieren, W., Kudritzki, R.: 2005, \journal{\apj}, \vol{634}, 1020.

Bresolin, F., Gieren, W., Kudritzki, R.P., Pietrzy{\'n}ski, G., Urbaneja, M.A., Carraro, G.: 2009, \journal{\apj}, \vol{700}, 309.

Butler, D.J., Mart{\'{\i}}nez-Delgado, D., Brandner, W.: 2004, \journal{\aj}, \vol{127}, 1472.

Cao, C., Wu, H.: 2007, \journal{\aj}, \vol{133}, 1710.

Cioffi, D.: 1990, in Brinkmann, W., Fabian, A., Giovannelli, F., eds, \journal{Physical Processes in Hot Cosmic Plasmas}, Kluwer Academic Publishers, Boston, 1.

Dai, H., Wang, T.: 2008, \journal{\cjaap}, \vol{8}, 245.

Dalcanton, J.J., Williams, B.F., Seth, A.C., Dolphin, A., Holtzman, J., Rosema, K., Skillman, E.D., Cole, A., Girardi, L., Gogarten, S.M., Karachentsev, I.D., Olsen, K., Weisz, D., Christensen, C., Freeman, K., Gilbert, K., Gallart, C., Harris, J., Hodge, P., de Jong, R.S., Karachentseva, V., Mateo, M., Stetson, P.B., Tavarez, M., Zaritsky, D., Governato, F., Quinn, T.: 2009, \journal{\apjs}, \vol{183}, 67.

de Grijs, R., Wilkinson, M.I., Tadhunter, C.N.: 2005, \journal{\mnras}, \vol{361}, 311.

de Mello, D.F., Smith, L.J., Sabbi, E., Gallagher, J.S., Mountain, M., Harbeck, D.R.: 2008, \journal{\aj}, \vol{135}, 548.

de Vaucouleurs, G., de Vaucouleurs, A., Corwin, H.G. Jr., Buta, R.J., Paturel, G., Fouque, P.: 1991, \journal{Third Reference Catalogue of Bright Galaxies}, Springer, New York.

Deharveng, L., Caplan, J., Lequeux, J., Azzopardi, M., Breysacher, J., Tarenghi, M., Westerlund, B.: 1988, \journal{\aaps}, \vol{73}, 407.

Dodorico, S., Benvenuti, P., Sabbadin, F.: 1978, \journal{\aap}, \vol{63}, 63.

Dodorico, S., Dopita, M.A., Benvenuti, P.: 1980, \journal{\aaps}, \vol{40}, 67.

Duric, N.: 2000a, in Berkhuijsen, E.M., Beck, R., Walterbos, R.A.M., eds, \journal{Proceedings 232. WE-Heraeus Seminar}, 127.

Duric, N.: 2000b, in Berkhuijsen, E.M., Beck, R., Walterbos, R.A.M., eds, \journal{Proceedings 232. WE-Heraeus Seminar}, 179.

Fesen, R.A., Hurford, A.P.: 1996, \journal{\apjs}, \vol{106}, 563.

Fesen, R.A., Blair, W.P., Kirshner, R.P.: 1985, \journal{\apj}, \vol{292}, 29.

Filipovi{\'c}, M.D., Haberl, F., Winkler, P.F., Pietsch, W., Payne, J.L., Crawford, E.J., de Horta, A.Y., Stootman, F.H., Reaser, B.E.: 2008, \journal{\aap}, \vol{485}, 63.

Filipovi{\'c}, M.D., Payne, J.L., Reid, W., Danforth, C.W., Staveley-Smith, L., Jones, P.A., White, G.L.: 2005, \journal{\mnras}, \vol{364}, 217.

Freedman, W.L., Madore, B.F., Hawley, S.L., Horowitz, I.K., Mould, J., Navarrete, M., Sallmen, S.: 1992, \journal{\apj}, \vol{396}, 80.

Freedman, W.L., Madore, B.F., Gibson, B.K., Ferrarese, L., Kelson, D.D., Sakai, S., Mould, J.R., Kennicutt, R.C. Jr., Ford, H.C.,  Graham, J.A., Huchra, J.P., Hughes, S.M.G., Illingworth, G.D., Macri, L.M., Stetson, P.B.: 2001, \journal{\apj}, \vol{553}, 47.

Ganda, K., Falc{\'o}n-Barroso, J., Peletier, R.F., Cappellari, M., Emsellem, E., McDermid, R.M., de Zeeuw, P.T., Carollo, C.M.: 2006, \journal{\mnras}, \vol{367}, 46.

Ghosh, K.K., Saripalli, L., Gandhi, P., Foellmi, C., Guti{\'e}rrez, C.M., L{\'o}pez-Corredoira, M.: 2009, \journal{\aj}, \vol{137}, 3263.

Gieren, W., Pietrzy{\'n}ski, G., Walker, A., Bresolin, F., Minniti, D., Kudritzki, R., Udalski, A., Soszy{\'n}ski, I., Fouqu{\'e}, P., Storm, J., Bono, G.: 2004, \journal{\aj}, \vol{128}, 1167.

Gieren, W., Pietrzy{\'n}ski, G., Soszy{\'n}ski, I., Bresolin, F., Kudritzki, R.P., Minniti, D., Storm, J.: 2005, \journal{\apj}, \vol{628}, 695.

Gil de Paz, A., Boissier, S., Madore, B.F., Seibert, M., Joe, Y.H., Boselli, A., Wyder, T.K., Thilker, D., Bianchi, L., Rey, S., Rich, R.M.,  Barlow, T.A., Conrow, T., Forster, K., Friedman, P.G., Martin, D.C., Morrissey, P., Neff, S.G., Schiminovich, D., Small, T., Donas, J., Heckman, T.M., Lee, Y., Milliard, B., Szalay, A.S., Yi, S.: 2007, \journal{\apjs}, \vol{173}, 185.

Girardi, L., Dalcanton, J., Williams, B., de Jong, R., Gallart, C., Monelli, M., Groenewegen, M.A.T., Holtzman, J.A., Olsen, K.A.G., Seth, A.C., Weisz, D.R., the ANGST/ANGRRR Collaboration: 2008, \journal{\pasp}, \vol{120}, 583.

Gliozzi, M., Satyapal, S., Eracleous, M., Titarchuk, L., Cheung, C.C.: 2009, \journal{\apj}, \vol{700}, 1759.

Gogarten, S.M., Dalcanton, J.J., Williams, B.F., Seth, A.C., Dolphin, A., Weisz, D., Skillman, E., Holtzman, J., Cole, A., Girardi, L., de Jong, R.S., Karachentsev, I.D., Olsen, K., Rosema, K.: 2009a, \journal{\apj}, \vol{691}, 115.

Gogarten, S.M., Dalcanton, J.J., Murphy, J.W., Williams, B.F., Gilbert, K., Dolphin, A.: 2009b, \journal{\apj}, \vol{703}, 300.

Gogarten, S.M., Dalcanton, J.J., Williams, B.F., Ro{\v s}kar, R., Holtzman, J., Seth, A.C., Dolphin, A., Weisz, D., Cole, A., Debattista, V.P., Gilbert, K.M., Olsen, K., Skillman, E., de Jong, R.S., Karachentsev, I.D., Quinn, T.R.: 2010, \journal{\apj}, \vol{712}, 858.

Gonz{\'a}lez Delgado, R.M., P{\'e}rez, E., Cid Fernandes, R., Schmitt, H.: 2008, \journal{\aj}, \vol{135}, 747.

Graham, A.W., Erwin, P., Caon, N., Trujillo, I.: 2001, \journal{\apjl}, \vol{563}, 11.

Guerrero, M.A., Chu, Y.: 2008, \journal{\apjs}, \vol{177}, 216.

Holwerda, B.W., Keel, W.C., Williams, B., Dalcanton, J.J., de Jong, R.S.: 2009, \journal{\aj}, \vol{137}, 3000.

Hoopes, C.G., Walterbos, R.A.M., Greenwalt, B.E.: 1996, \journal{\aj}, \vol{112}, 1429.

Karachentsev, I.D., Grebel, E.K., Sharina, M.E., Dolphin, A.E., Geisler, D., Guhathakurta, P., Hodge, P.W., Karachentseva, V.E., Sarajedini, A., Seitzer, P.: 2003, \journal{\aap}, \vol{404}, 93.

Kim, S.C., Sung, H., Park, H.S., Sung, E.C.: 2004, \journal{\cjaap}, \vol{4}, 299.

Kornei, K.A., McCrady, N.: 2009, \journal{\apj}, \vol{697}, 1180.

Kudritzki, R., Urbaneja, M.A., Bresolin, F., Przybilla, N., Gieren, W., Pietrzy{\'n}ski, G.: 2008, \journal{\apj}, \vol{681}, 269.

Kuntz, K.D., Snowden, S.L.: 2010, \journal{\apjs}, \vol{188}, 46.

Larsen, S.S.: 2004, \journal{\aap}, \vol{416}, 537.

Lauer, T.R., Faber, S.M., Gebhardt, K., Richstone, D., Tremaine, S., Ajhar, E.A., Aller, M.C., Bender, R., Dressler, A., Filippenko, A.V., Green, R., Grillmair, C.J., Ho, L.C., Kormendy, J., Magorrian, J., Pinkney, J., Siopis, C.: 2005, \journal{\aj}, \vol{129}, 2138.

Lauer, T.R., Gebhardt, K., Faber, S.M., Richstone, D., Tremaine, S., Kormendy, J., Aller, M.C., Bender, R., Dressler, A., Filippenko, A.V., Green, R., Ho, L.C.: 2007a, \journal{\apj}, \vol{664}, 226.

Lauer, T.R., Faber, S.M., Richstone, D., Gebhardt, K., Tremaine, S., Postman, M., Dressler, A., Aller, M.C., Filippenko, A.V., Green, R., Ho, L.C., Kormendy, J., Magorrian, J., Pinkney, J.: 2007b, \journal{\apj}, \vol{662}, 808.

Lianou, S., Grebel, E.K., Koch, A.: 2009, \journal{\an}, \vol{330}, 995.

Long, K.S.: 1996, in McCray, R., Wang, Z., eds, \journal{IAU Colloq. 145: Supernovae and Supernova Remnants}, Cambridge University Press, 1996, 349.

Mathewson, D.S., Clarke, J.N.: 1972, \journal{\apjl}, \vol{178}, 105.

Mathewson, D.S., Clarke, J.N.: 1973a, \journal{\apj}, \vol{179}, 89.

Mathewson, D.S., Clarke, J.N.: 1973b, \journal{\apj}, \vol{180}, 725.

Mathewson, D.S., Clarke, J.N.: 1973c, \journal{\apj}, \vol{182}, 697.

Matonick, D.M., Fesen, R.A.: 1997, \journal{\apjs}, \vol{112}, 49.

Maund, J.R., Smartt, S.J.: 2009, \journal{\sci}, \vol{324}, 486.

Melbourne, J., Williams, B., Dalcanton, J., Ammons, S.M., Max, C., Koo, D.C., Girardi, L., Dolphin, A.: 2010, \journal{\apj}, \vol{712}, 469.

Mickaelian, A.M.: 2004, \journal{\aap}, \vol{426}, 367.

Millar, W.C., White, G.L., Filipovi{\'c}, M.D., Payne, J.L., Crawford, E.J., Pannuti, T.G., Staggs, W.D.: 2011, \journal{\aaps}, \vol{332}, 221.

Milone, A.P., Villanova, S., Bedin, L.R., Piotto, G., Carraro, G., Anderson, J., King, I.R., Zaggia, S.: 2006, \journal{\aap}, \vol{456}, 517

Montes, G., P{\'e}rez-Torres, M.A., Alberdi, A., Gonz{\'a}lez, R.F.: 2009, \journal{\apj}, \vol{705}, 899.

Mould, J., Sakai, S.: 2008, \journal{\apjl}, \vol{686}, 75.

Nantais, J.B., Huchra, J.P., Barmby, P., Olsen, K.A.G.: 2010, \journal{\aj}, \vol{139}, 1178.

Pannuti, T.G., Duric, N., Lacey, C.K., Goss, W.M., Hoopes, C.G., Walterbos, R.A.M., Magnor, M.A.: 2000, \journal{\apj}, \vol{544}, 780.

Pannuti, T.G., Schlegel, E.M., Lacey, C.K.: 2007, \journal{\aj}, \vol{133}, 1361.

Payne, J.L., Filipovi{\'c}, M.D., Pannuti, T.G., Jones, P.A., Duric, N., White, G.L., Carpano, S.: 2004, \journal{\aap}, \vol{425}, 443.

Payne, J.L., White, G.L., Filipovi{\'c}, M.D.: 2008, \journal{\mnras}, \vol{383}, 1175.

Peeples, M.S., Martini, P.: 2006, \journal{\apj}, \vol{652}, 1097.

Pietrzy{\'n}ski, G., Gieren, W., Fouqu{\'e}, P., Pont, F.: 2001, \journal{\aap}, \vol{371}, 497.

Puche, D., Carignan, C., Bosma, A.: 1990, \journal{\aj}, \vol{100}, 1468.

Read, A.M., Pietsch, W.: 2001, \journal{\aap}, \vol{373}, 473.

Read, A.M., Ponman, T.J., Strickland, D.K.: 1997, \journal{\mnras}, \vol{286}, 626.

Richstone, D.: 2000, Hubble Space Telescope Proposal 8591.

Rizzi, L., Bresolin, F., Kudritzki, R., Gieren, W., Pietrzy{\'n}ski, G.: 2006, \journal{\apj}, \vol{638}, 766.

Rizzi, L., Tully, R.B., Makarov, D., Makarova, L., Dolphin, A.E., Sakai, S., Shaya, E.J.: 2007, \journal{\apj}, \vol{661}, 815.

Rosolowsky, E., Blitz, L.: 2005, \journal{\apj}, \vol{623}, 826.

Rossa, J., van der Marel, R.P., B{\"o}ker, T., Gerssen, J., Ho, L.C., Rix, H., Shields, J.C., Walcher, C.: 2006, \journal{\aj}, \vol{132}, 1074.

Schild, H., Crowther, P.A., Abbott, J.B., Schmutz, W.: 2003, \journal{\aap}, \vol{397}, 859.

Schinnerer, E., B{\"o}ker, T., Meier, D.S.: 2003, \journal{\apjl}, \vol{591}, 115.

Schinnerer, E., B{\"o}ker, T., Emsellem, E., Lisenfeld, U.: 2006, \journal{\apj}, \vol{649}, 181.

Seth, A., Ag{\"u}eros, M., Lee, D., Basu-Zych, A.: 2008, \journal{\apj}, \vol{678}, 116.

Shaw, R.A., Reid, W.A., Parker, Q.A.: 2007, \journal{\pasp}, \vol{119}, 19.

Shaw, R.A., Stanghellini, L., Villaver, E., Mutchler, M.: 2006, \journal{\apjs}, \vol{167}, 201.

Siopis, C., Gebhardt, K., Lauer, T.R., Kormendy, J., Pinkney, J., Richstone, D., Faber, S.M., Tremaine, S., Aller, M.C., Bender, R., Bower,  G., Dressler, A., Filippenko, A.V., Green, R., Ho, L.C., Magorrian, J.: 2009, \journal{\apj}, \vol{693}, 946.

Smartt, S.J., Maund, J.R., Hendry, M.A., Tout, C.A., Gilmore, G.F., Mattila, S., Benn, C.R.: 2004, \journal{\sci}, \vol{303}, 499.

Soffner, T., Mendez, R.H., Jacoby, G.H., Ciardullo, R., Roth, M.M., Kudritzki, R.P.: 1996, \journal{\aap}, \vol{306}, 9.

Strom, R.C.: 1996, in McCray, R., Wang, Z., eds, \journal{IAU Colloq. 145: Supernovae and Supernova Remnants}, Cambridge University Press, 1996, 333.

Sugerman, B.E.K.: 2005, \journal{\apjl}, \vol{632}, 17.

Sugerman, B.E.K., Ercolano, B., Barlow, M.J., Tielens, A.G.G.M., Clayton, G.C., Zijlstra, A.A., Meixner, M., Speck, A., Gledhill, T.M.,  Panagia, N., Cohen, M., Gordon, K.D., Meyer, M., Fabbri, J., Bowey, J.E., Welch, D.L., Regan, M.W., Kennicutt, R.C.: 2006, \journal{\sci}, \vol{313}, 196.

Tikhonov, N.A., Galazutdinova, O.A.: 2005a, \journal{\asph}, \vol{48}, 221.

Tikhonov, N.A., Galazutdinova, O.A.: 2009, \journal{\al}, \vol{35}, 748.

Tikhonov, N.A., Galazutdinova, O.A., Drozdovsky, I.O.: 2005b, \journal{\aap}, \vol{431}, 127.

Tully, R.B., Fisher, J.R.: 1988, \journal{Catalog of Nearby Galaxies}, Cambridge University Press, 1988.

Tully, R.B., Rizzi, L., Dolphin, A.E., Karachentsev, I.D., Karachentseva, V.E., Makarov, D.I., Makarova, L., Sakai, S., Shaya, E.J.: 2006, \journal{\aj}, \vol{132}, 729.

Uro{\v s}evi{\'c}, D., Pannuti, T.G., Duric, N., Theodorou, A.: 2005, \journal{\aap}, \vol{435}, 437.

Van Dyk, S.D., Li, W., Filippenko, A.V.: 2003, \journal{\pasp}, \vol{115}, 1289.

Wadadekar, Y., Casertano, S., Hook, R., K{\i}z{\i}ltan, B., Koekemoer, A., Ferguson, H., Denchev, D.: 2006, \journal{\pasp}, \vol{118}, 450.

Walcher, C.J., van der Marel, R.P., McLaughlin, D., Rix, H., B{\"o}ker, T., H{\"a}ring, N., Ho, L.C., Sarzi, M., Shields, J.C.: 2005, \journal{\apj}, \vol{618}, 237.

Williams, B.F., Dalcanton, J.J., Seth, A.C., Weisz, D., Dolphin, A., Skillman, E., Harris, J., Holtzman, J., Girardi, L., de Jong, R.S., Olsen, K., Cole, A., Gallart, C., Gogarten, S.M., Hidalgo, S.L., Mateo, M., Rosema, K., Stetson, P.B., Quinn, T.: 2009, \journal{\aj}, \vol{137}, 419.

Williams, B.F., Dalcanton, J.J., Dolphin, A.E., Holtzman, J., Sarajedini, A.: 2009, \journal{\apjl}, \vol{695}, 15.

Williams, B.F., Dalcanton, J.J., Stilp, A., Gilbert, K.M., Ro{\v s}kar, R., Seth, A.C., Weisz, D., Dolphin, A., Gogarten, S.M., Skillman, E., Holtzman, J.: 2010, \journal{\apj}, \vol{709}, 135.

Windhorst, R.A., Taylor, V.A., Jansen, R.A., Odewahn, S.C., Chiarenza, C.A.T., Conselice, C.J., de Grijs, R., de Jong, R.S., MacKenty, J.W., Eskridge, P.B., Frogel, J.A., Gallagher, J.S. III, Hibbard, J.E., Matthews, L.D., O'Connell, R.W.: 2002, \journal{\apjs}, \vol{143}, 113.

\endreferences

%% file: Figures.tex

\newcommand\figure[4][]{\begin{centering}%
\includegraphics[#1]{#3}\par\vspace{3mm}
{\textbf{Figure #2}. #4}
\end{centering}}

\newpage

\figure{1}{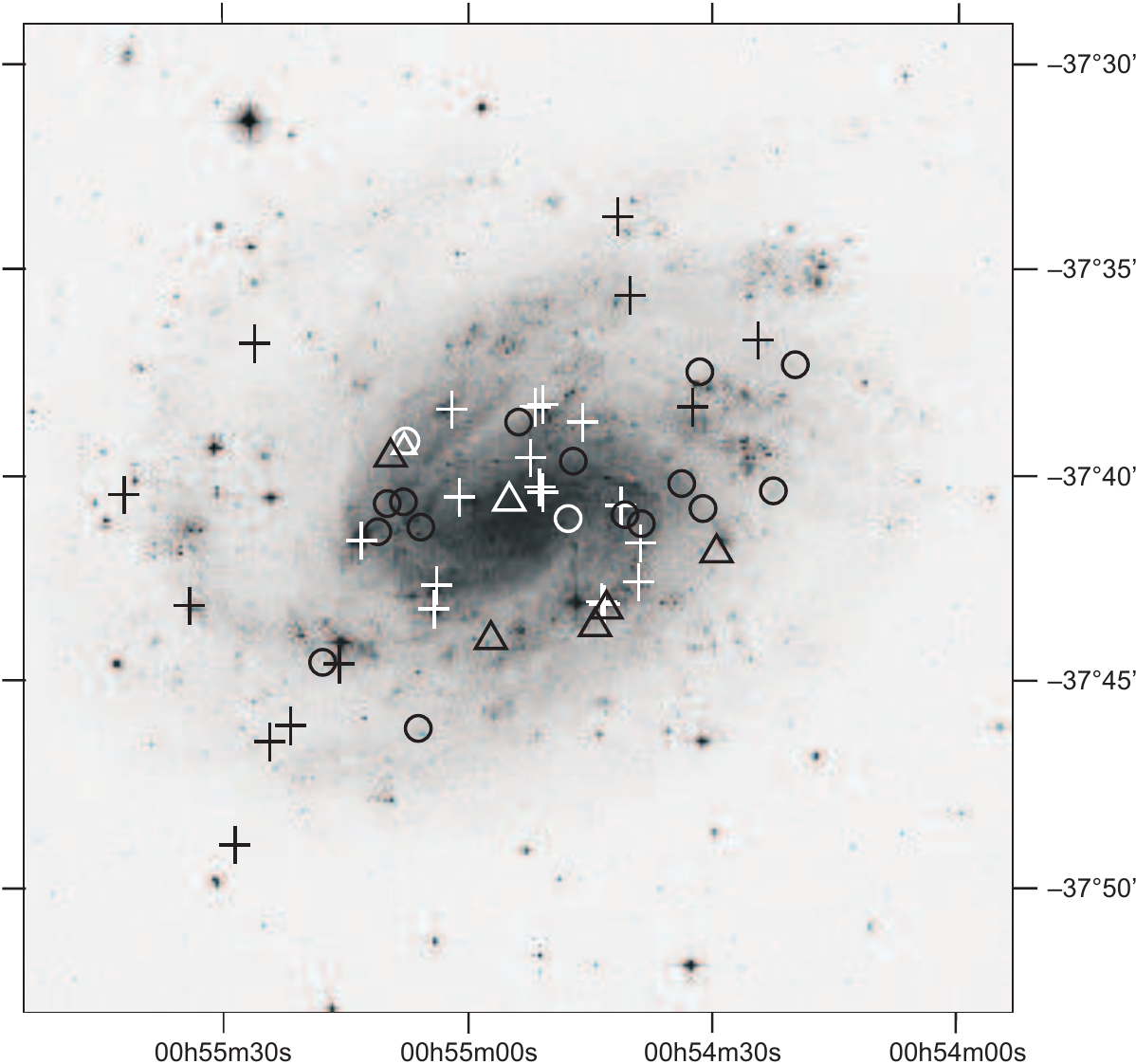}
{A DSS image of NGC\,300 with the positions (in J2000.0 coordinates) indicated of the 51 SNRs and candidate SNRs considered by the present study. Radio sources (SNRs and SNR candidates only) from P04 are shown with crosses. Optical candidates with line ratios measured with long-slit spectra (from BL97) are shown as circles and optical candidates with line ratios measured by interference filters (BL97) are shown with triangles. Symbols are black or white only for increased contrast. (Figure originally published in MWF11. Southern sky DSS image, courtesy of Royal Observatory Edinburgh, Anglo-Australian Observatory, California Institute of Technology.)}
%

\newpage

\figure[width=6.75in]{2}{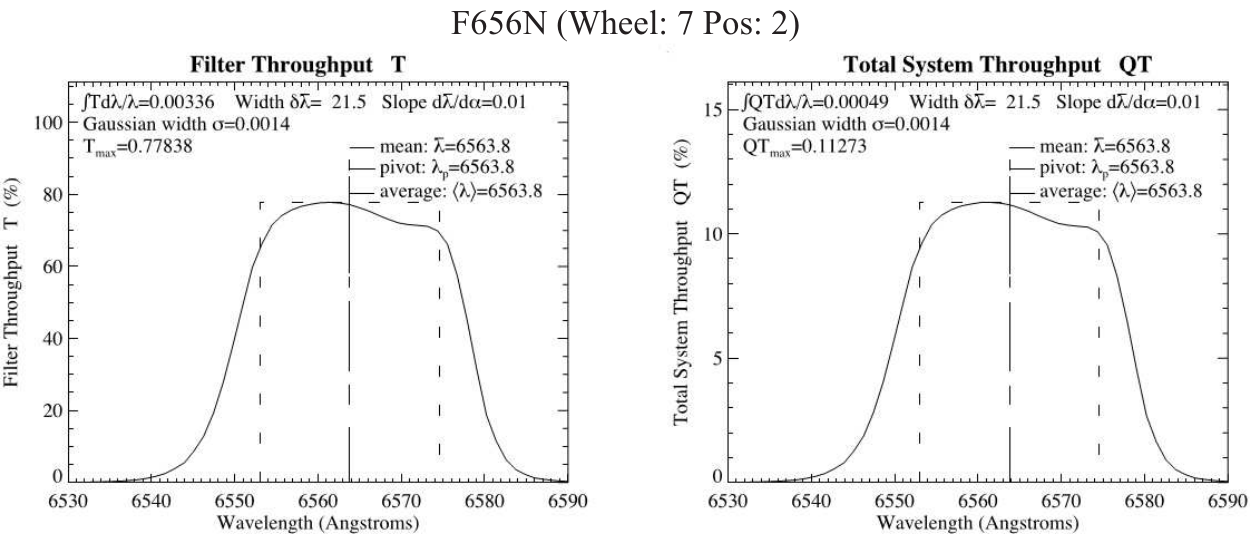}
{The pass band characteristics of the N565 filter used on the WFPC2 of the HST. (STScI Institute, WFPC2 Observer's Handbook.)}

\newpage

\figure{3}{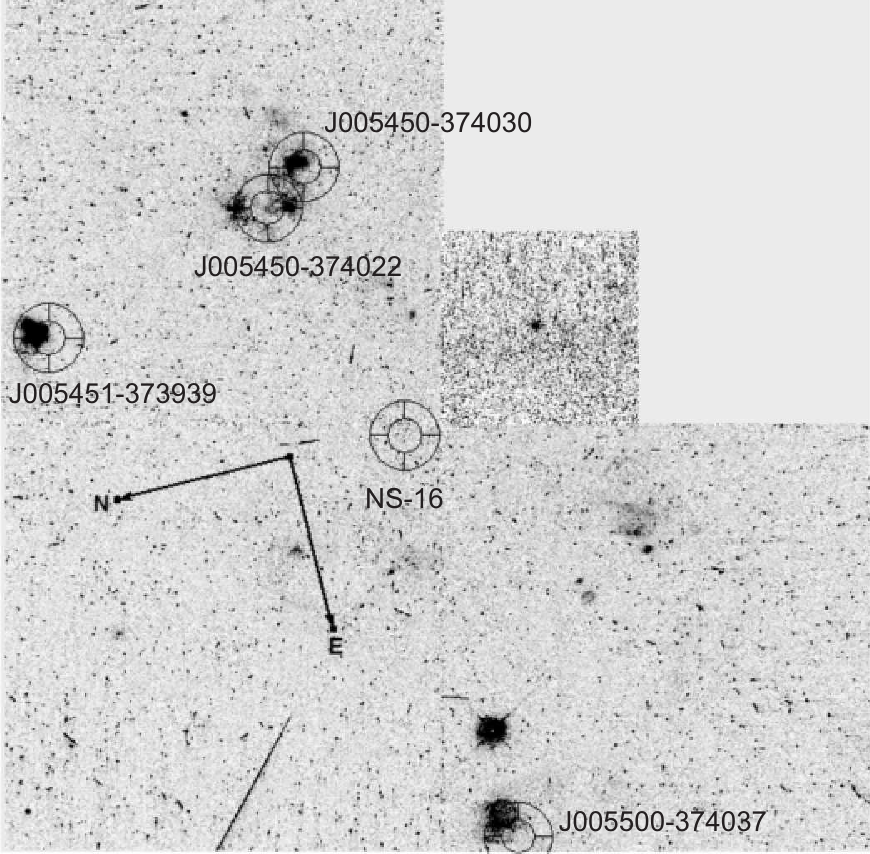}
{This image is contained in the file \texttt{u6713709r\_drz.fits} and is here modified using the SAO's DS9 software. The image is a 400 second exposure centered on $\alpha = 00$ 54 54.54, $\delta = -37$ 40 35.9 and rotated \Sim\Deg{104} east of north. The high-resolution CCD (first quadrant) is centered on the nucleus of \TG. The five SNR candidates in this image are described in Table~6. The pandas are centered on each candidate with an inner circle diameter of \Sec{6} (61.2~pc).}

\newpage

\figure[width=4in]{4}{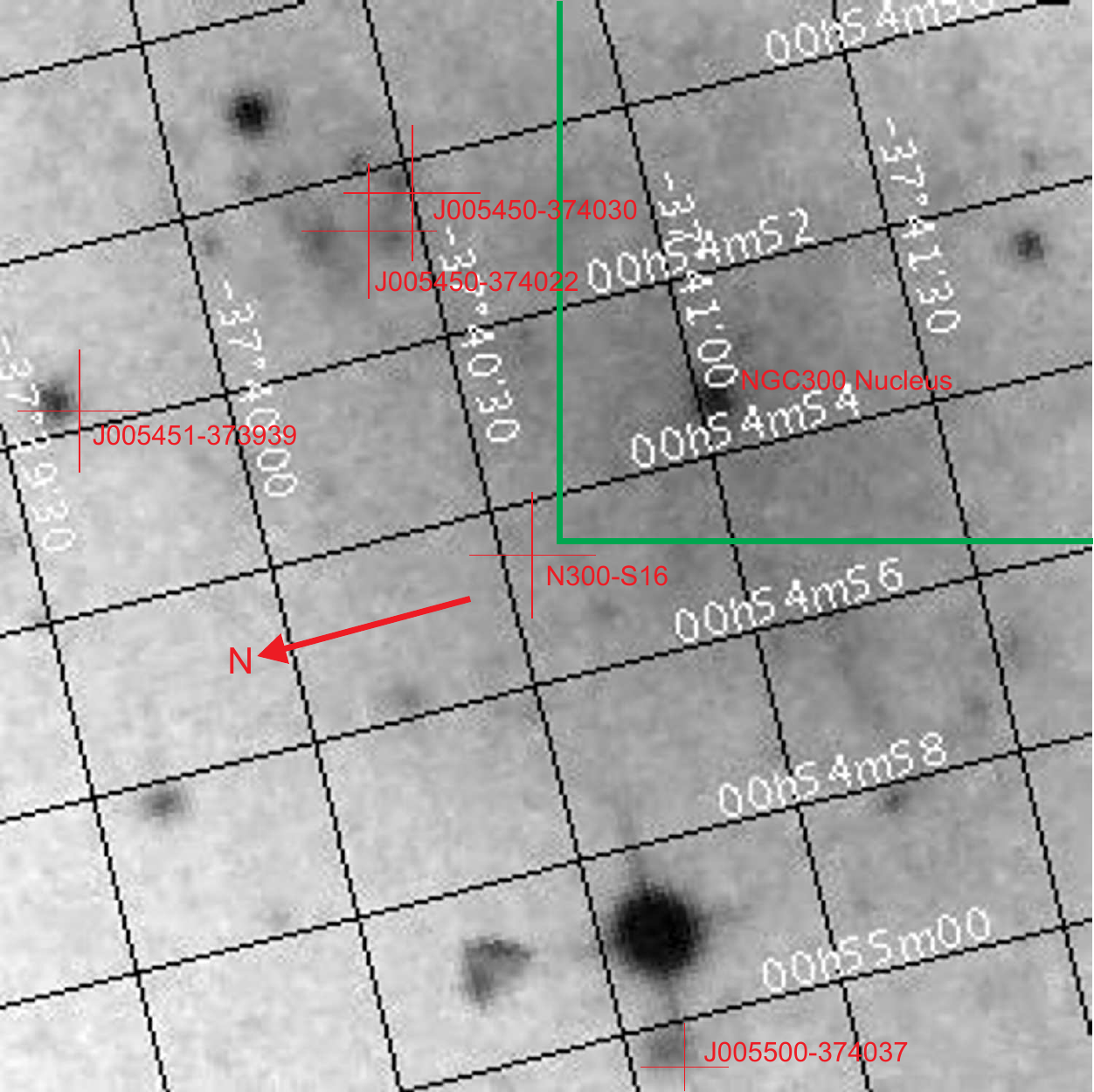}
{This is the approximate equivalent to the HST image of Fig.~3 created with SkyView using DSS2-Red data. (Southern sky DSS image, Royal Observatory Edinburgh, Anglo-Australian Observatory, California Institute of Technology.)}

\newpage

\figure{5}{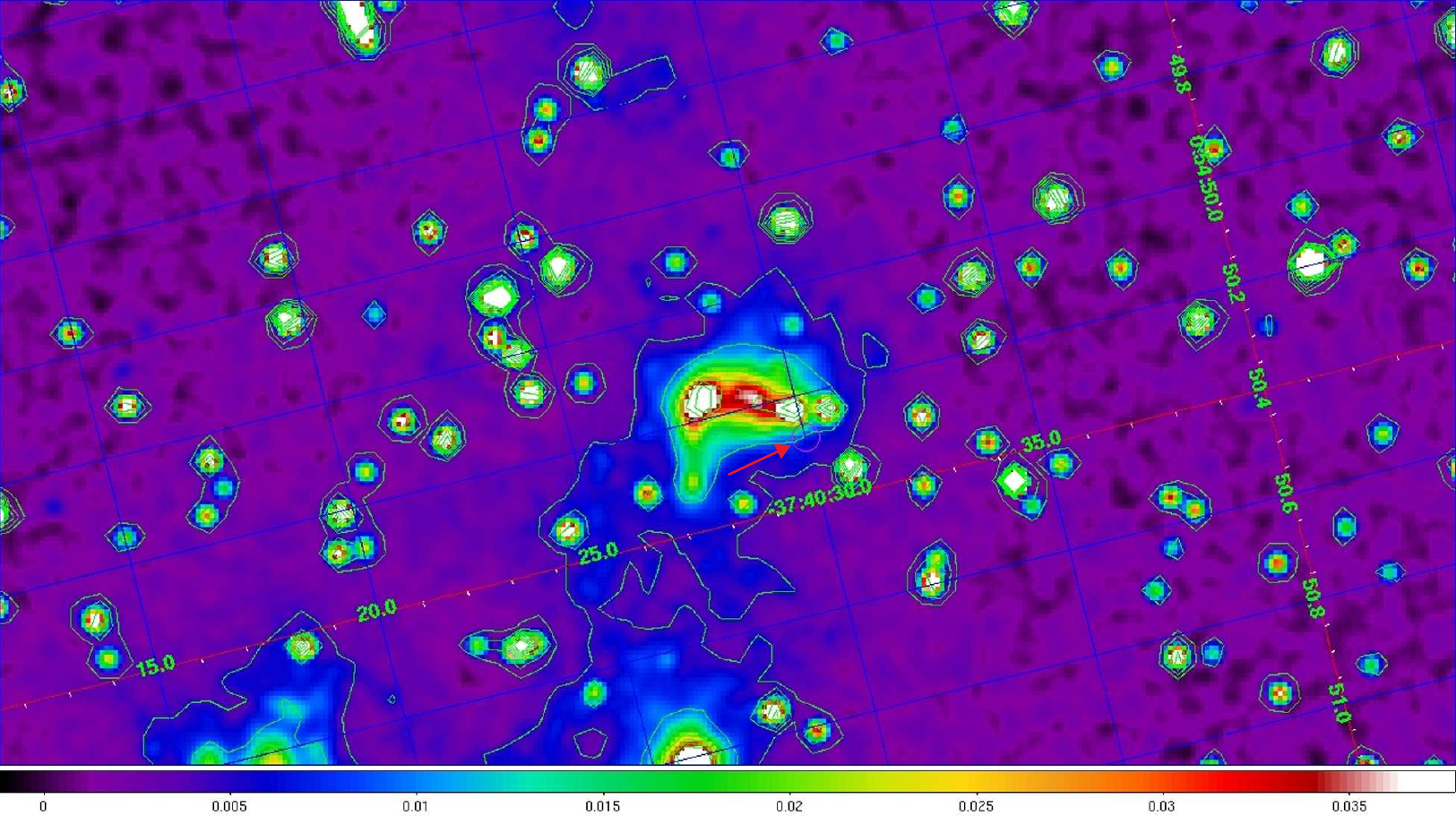}
{The region surrounding the P04 candidate \ATCA{5450}{4030}. The candidate location is shown by the magenta circle (red arrow pointing to it) just to the lower right of center of the intense \Hii\ emission. The circle is \Sec{1} (10.2~pc) in diameter -- about twice the $1\sigma$ positional error. This false color image and contour plot was created with DS9. Zoom level: 4; Scale: Linear, 98\%; Color: SLS; WCS: Equatorial, J2000; Analysis: Contours: 8, Smooth Parameters: 3, Gaussian.}

\newpage

\figure{6}{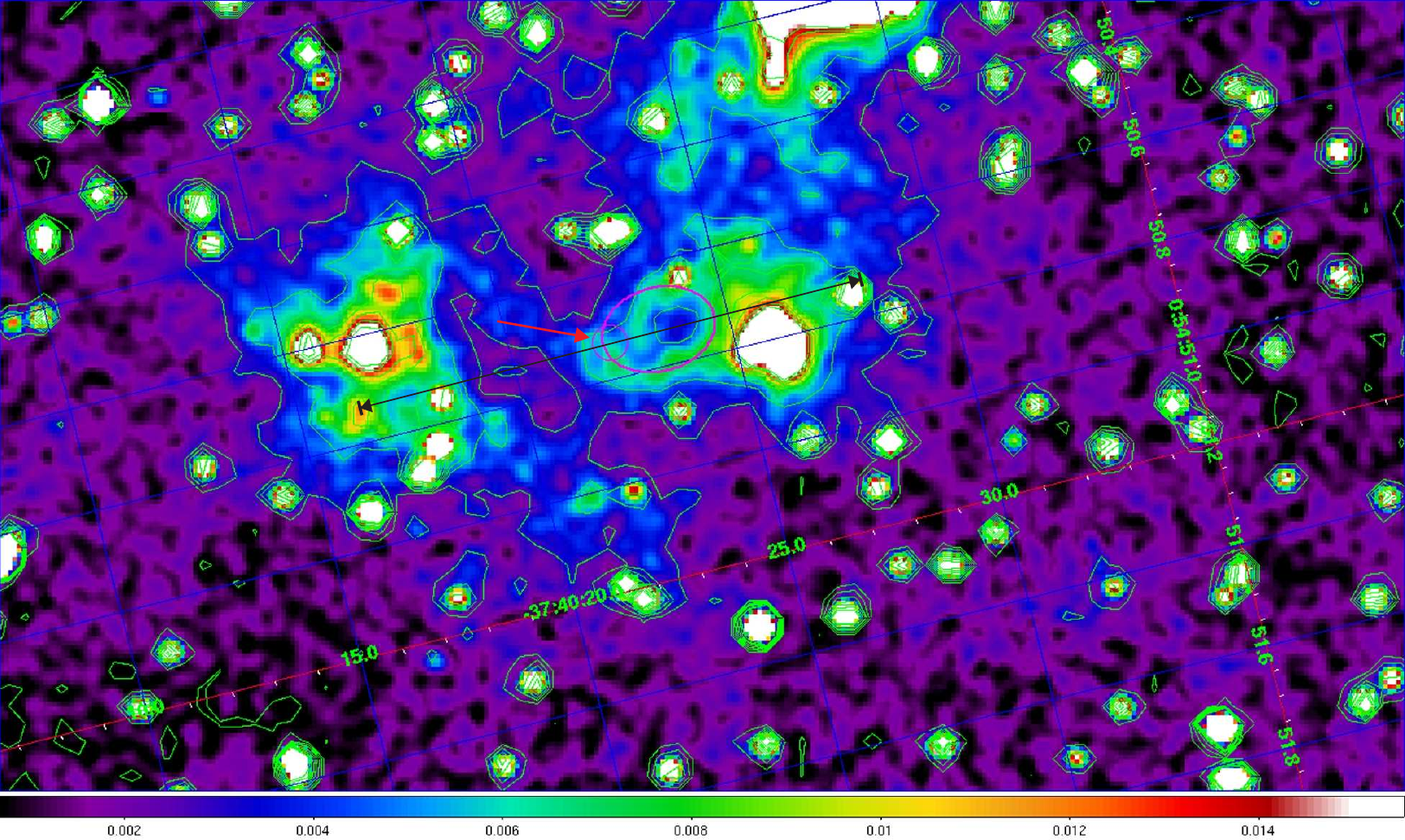}
{The region surrounding the P04 candidate \ATCA{5450}{4022}. The radio source location is shown by the magenta circle (red arrow pointing to it). The circle is \Sec{1} in diameter (\Sim{2$\sigma$} positional error). This false color image and contour plot was created with DS9. Zoom level: 4; Scale: Linear, 98\%; Color: SLS; WCS: Equatorial, J2000; Analysis: Contours: 8, Smooth Parameters: 3, Gaussian.}

\newpage

\figure{7}{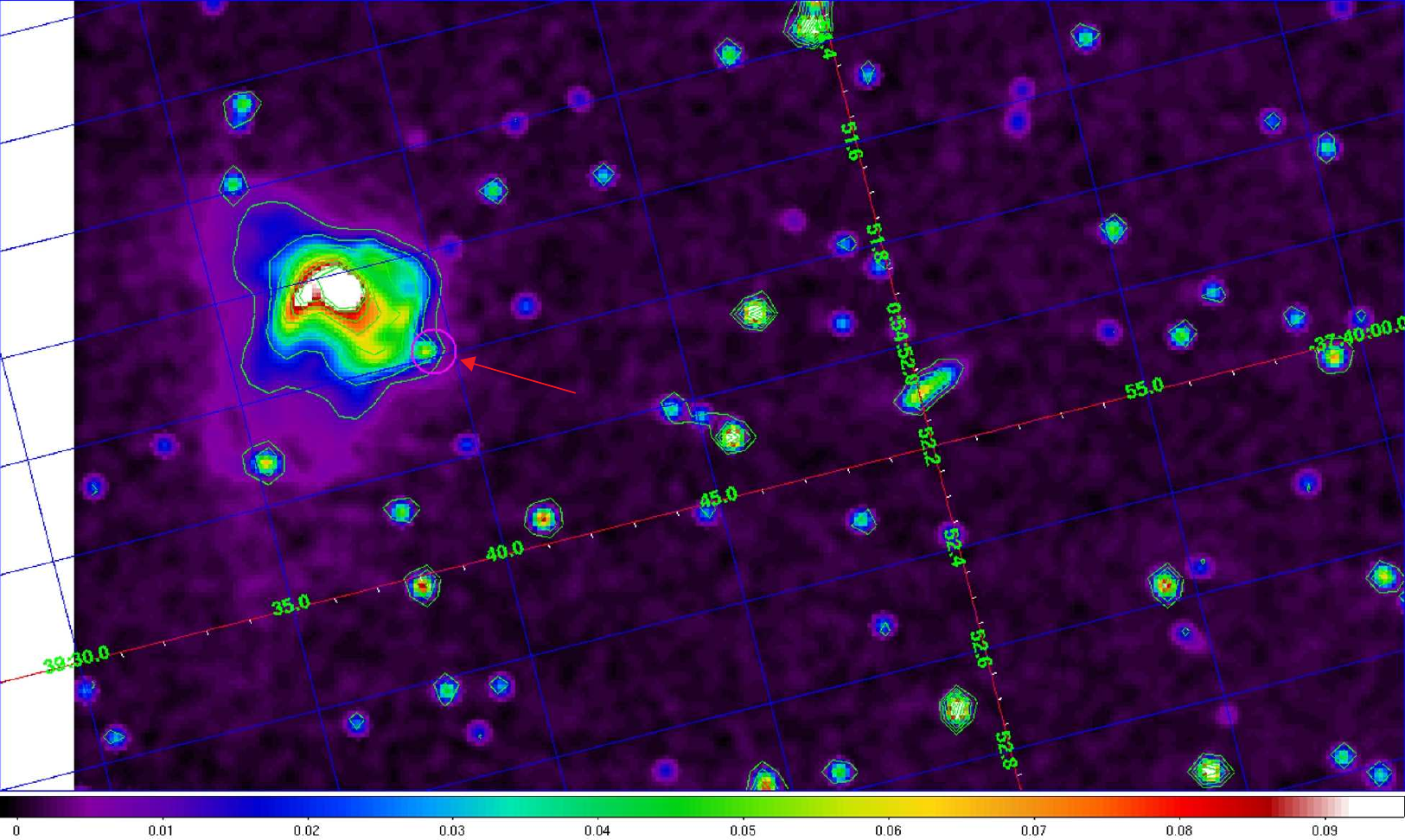}
{The region surrounding the P04 candidate \ATCA{5451}{3939}. The radio source location is shown by the magenta circle (red arrow pointing to it) \Sec{1} (10.2~pc) in diameter (\Sim{$2\sigma$} positional error). This false color image and contour plot was created with DS9. Zoom level: 4; Scale: Linear, 99.5\%; Color: SLS; WCS: Equatorial, J2000; Analysis: Contours: 10, Smooth Parameters: 3, Gaussian.}

\newpage

\figure{8}{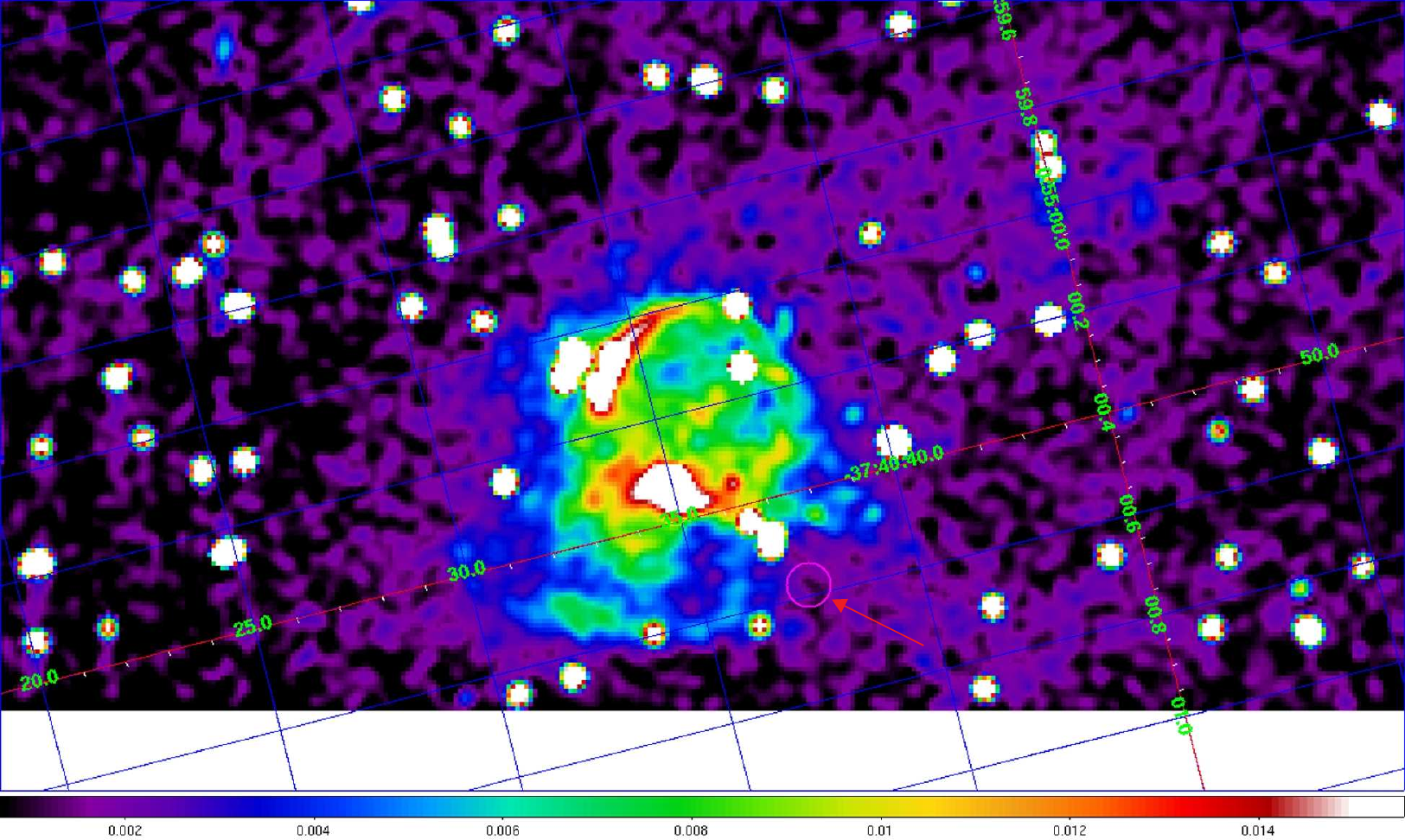}
{The region surrounding the P04 candidate \ATCA{5500}{4037}. The radio source location is shown by the magenta circle (red arrow pointing to it) \Sec{1} in diameter (\Sim{$2\sigma$} positional error). This false color image was created with DS9. Zoom level: 4; Scale: Linear, 95\%; Color: SLS; WCS: Equatorial, J2000. The contours were not used on this image.}

\vspace{1cm}

\figure{9}{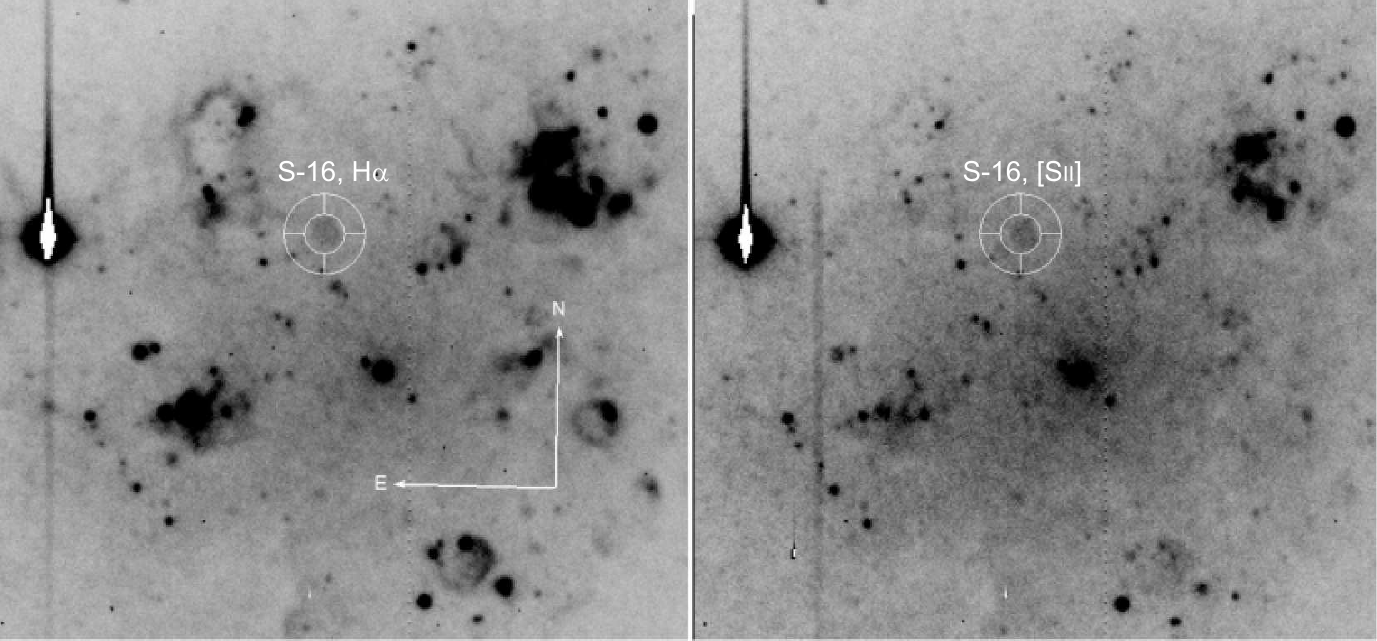}
{The left image is the \Ha\ filter image from BL97. The right image is the \Sii\ filter image of BL97. An object which could be \NS16 is visible as a faint smudge in these images. The inner circle of the panda is \Sec{8} (81.6~pc) in diameter.}

\newpage

\figure{10}{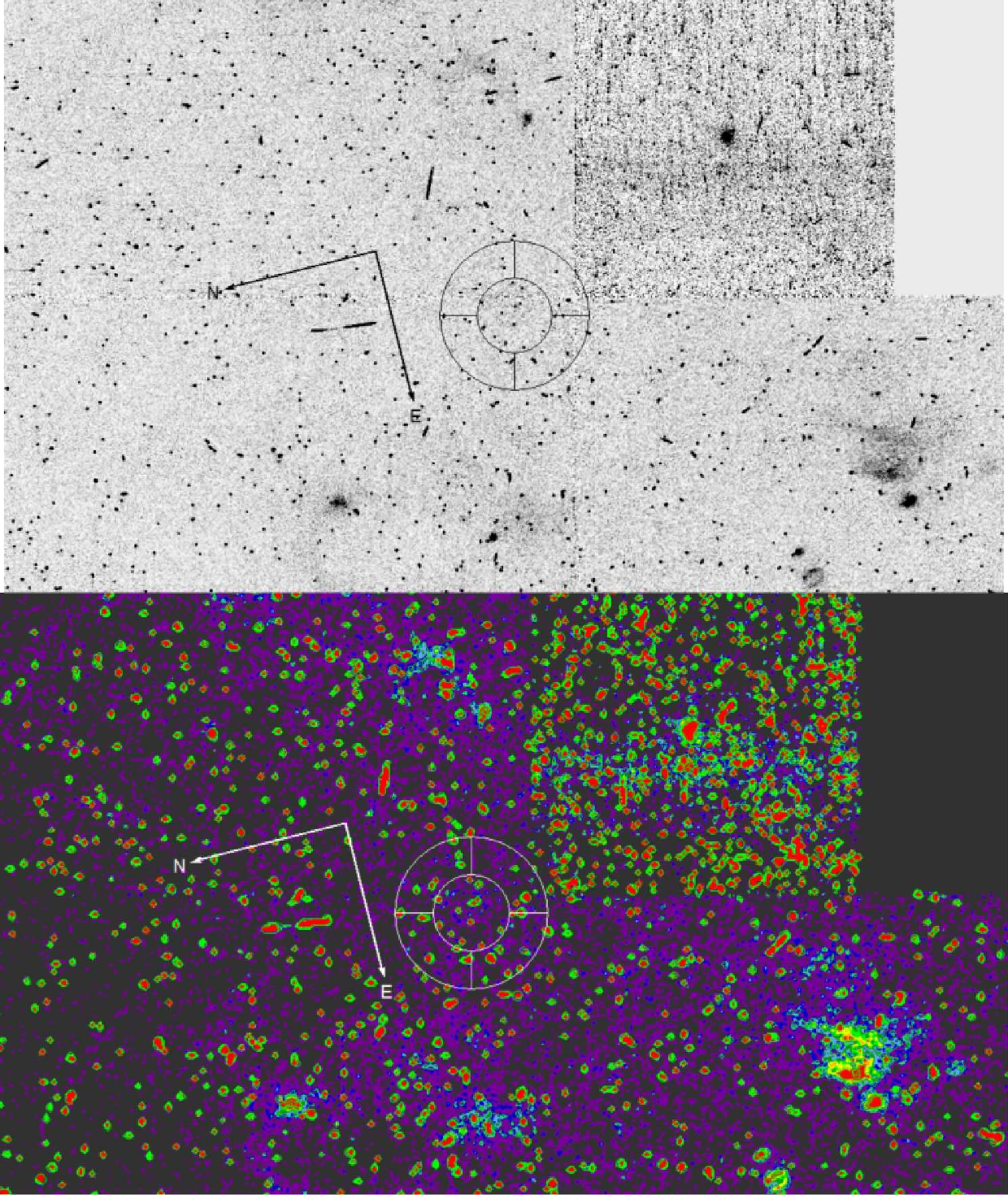}
{The HST \Ha\ image of \NS16 from shown in inverted gray-scale on top and in AIPS0 with contours on the bottom. In either case no discernible structure is seen in the H gas at the reported location of this candidate SNR. The panda inner circle is \Sec{8} (81.6~pc) in diameter.}

\newpage

\figure[width=6.25in]{11}{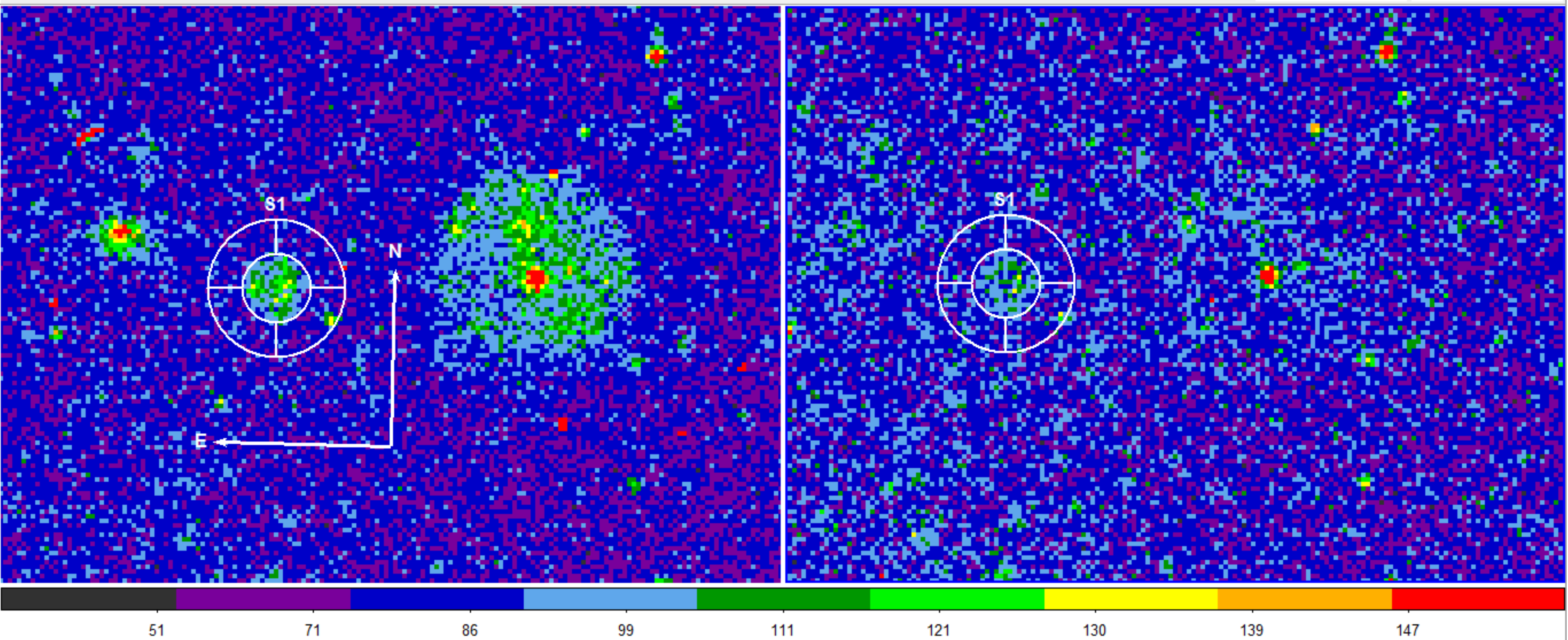}
{This is the BL97 published location of \NS1 ($\alpha$ 00 54 19.21, $\delta$ $-37$ 37 23.96) as found on the BL97 images (Image D) with 2MASS calibrated astrometry. The \Ha\ image is on the left and the \Sii\ image is on the right. The pandas on both images are centered on the published coordinates for \NS16, as guided by the 2MASS calibrated WCS for these images. The inner circle of the panda is \Sec{5.1} in diameter (52~pc) and the outer circle is \Sec{10.2} (104~pc) in diameter. The color scale is AIPS0, Zoom = 4, Scale = ZScale, Squared.}

\vspace{1cm}

\figure[width=6in]{12}{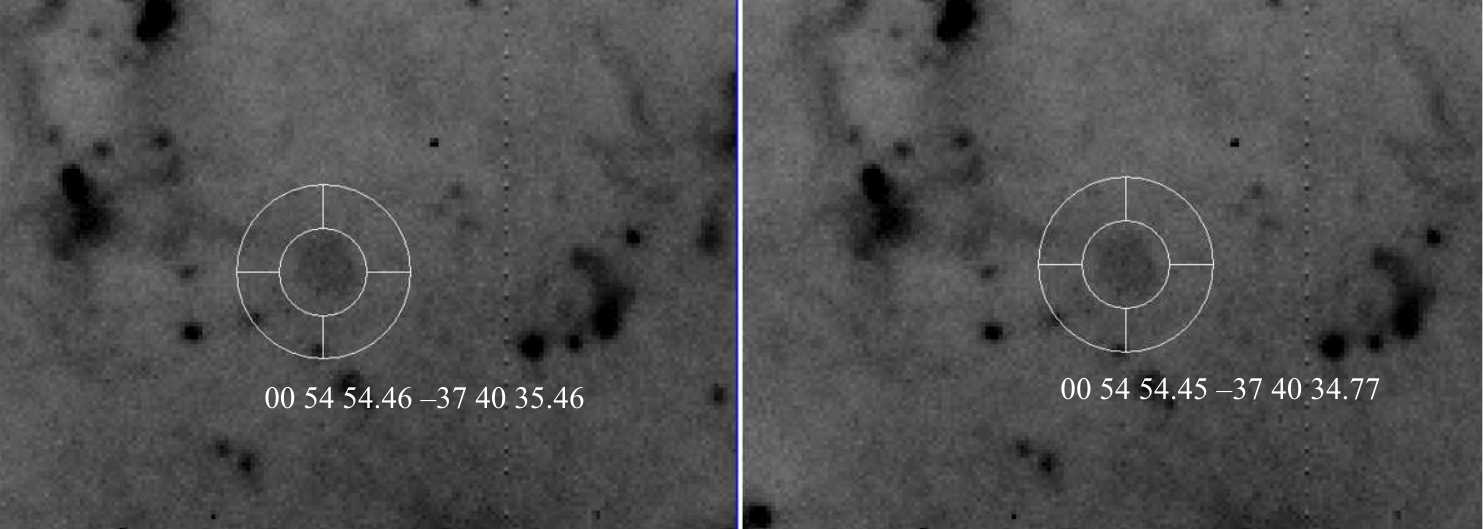}
{This is the BL97 location of \NS16. The \Ha\ G image is shown on both the left and right with north up and east to the left. The inner circle of the pandas is \Sec{4}. On the left is the reported position of NS16 (BL97) and on the right is the apparent center of the assumed (probable) image of the SNR candidate on the image. The astrometry error is within the typical seeing of two-meter class telescopes and within the reported seeing conditions for the observation.}

\newpage

\figure[width=6in]{13}{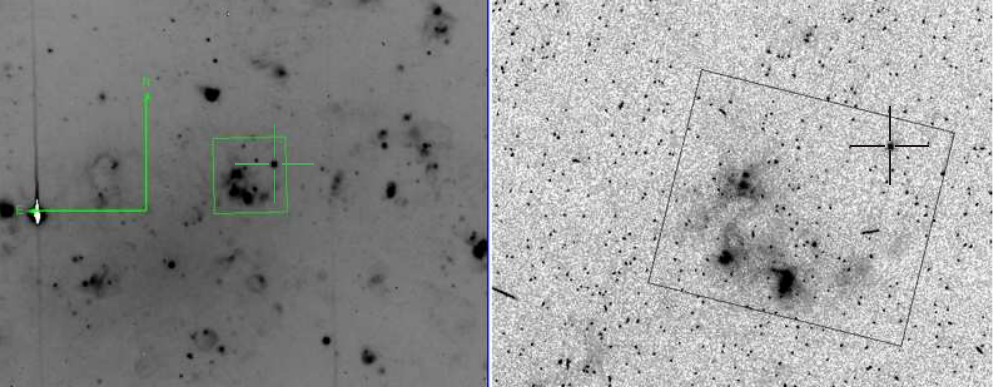}
{A comparison of the position of a bright star in BL97 G \Ha\ image and the HST image \texttt{u6713709r\_drz.fits}. The images are both rotated to show north as up, east to the left. The boxes show a set of \Hiir s used to help identify the star and the selected star is marked with a cross. The measured difference in position is shown in Table~9.}

\vspace{1cm}

\figure[width=6in]{14}{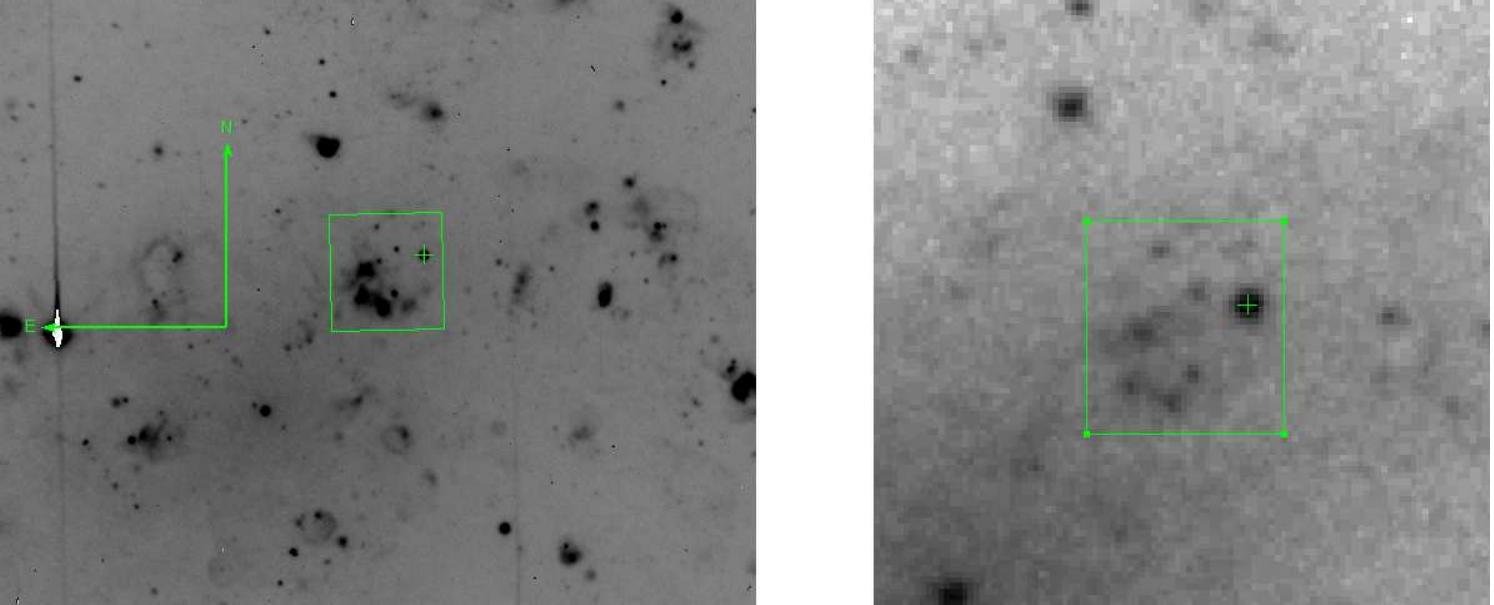}
{A comparison of the position of a bright star in BL97 G \Ha\ image and a DSS2-Red image containing the same \Hiir s. The image frames are both rotated to show north as up, east to the left. The boxes show a set of \Hiir s used to help identify the star and the selected star is marked with a small cross. The measured difference in position is shown in Table~9.}

\newpage

\figure[width=4in]{15}{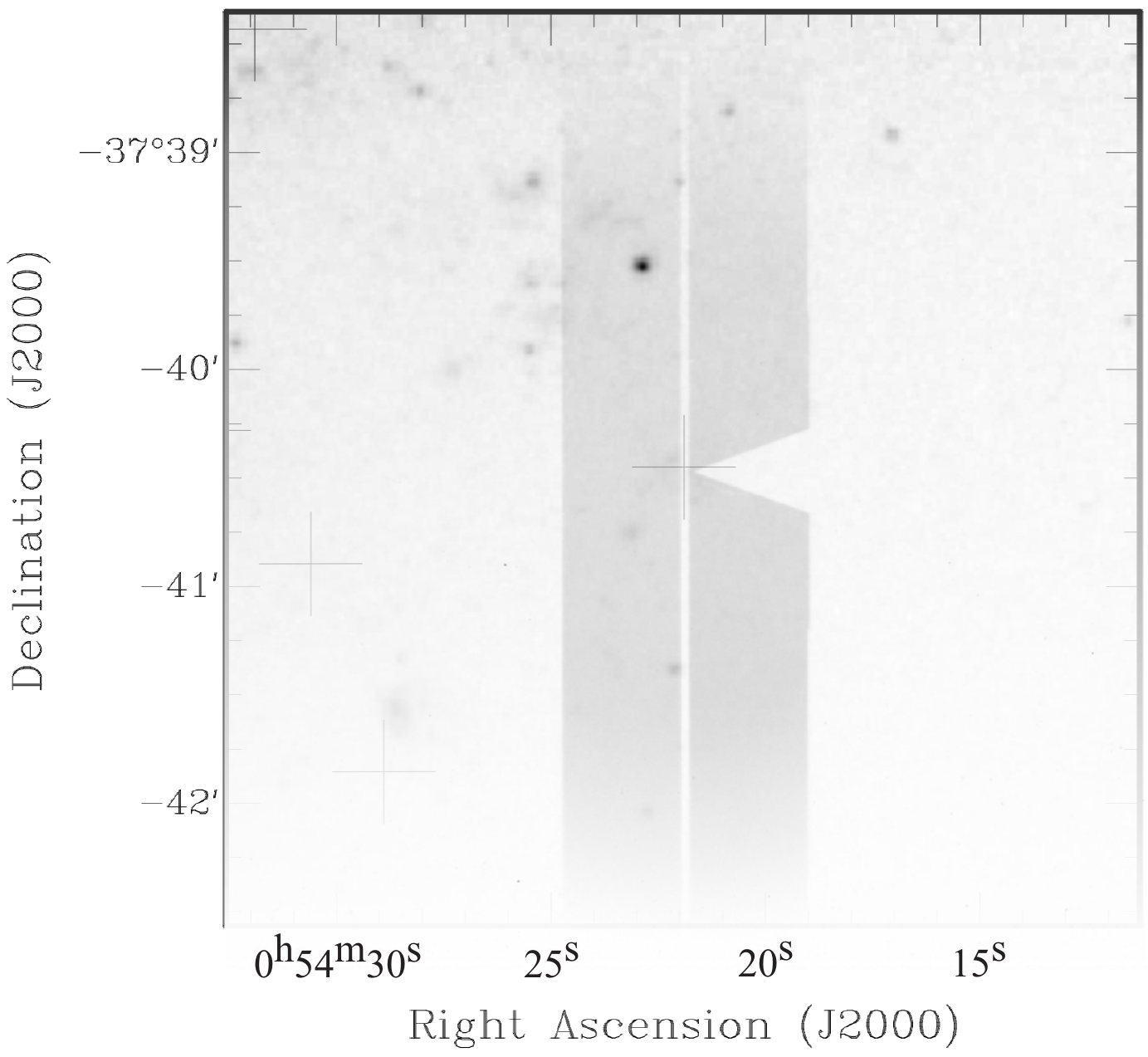}
{The DBS slit camera image laid on top of the MWF11 finding map for BL97 candidate \NS2. The circled objects are those which were used for the alignment of the two images. The pointing was exact but there is a significant difference between the measured \linrat: BL97 = 0.49 and MWF11 = \RatwErr{0.72}{0.39}. The large error in flux estimate also allows this object to be below the \linrat\ critical value.}

\newpage

\figure{16}{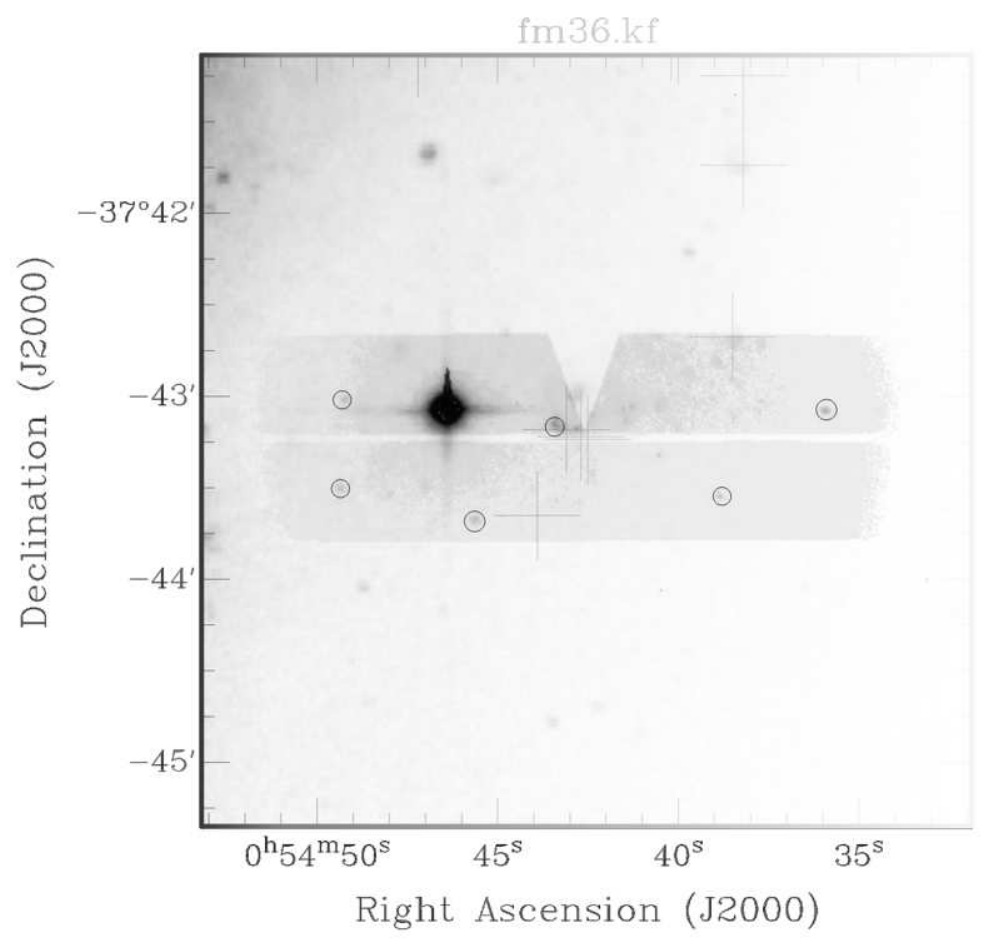}
{The DBS slit camera image laid on top of the MWF11 finding map for BL97 candidate \NS11 (lower right cross). The circled objects are those which were used for the alignment of the two images. In this case there is significant source confusion. See text.}

\newpage

\figure{17}{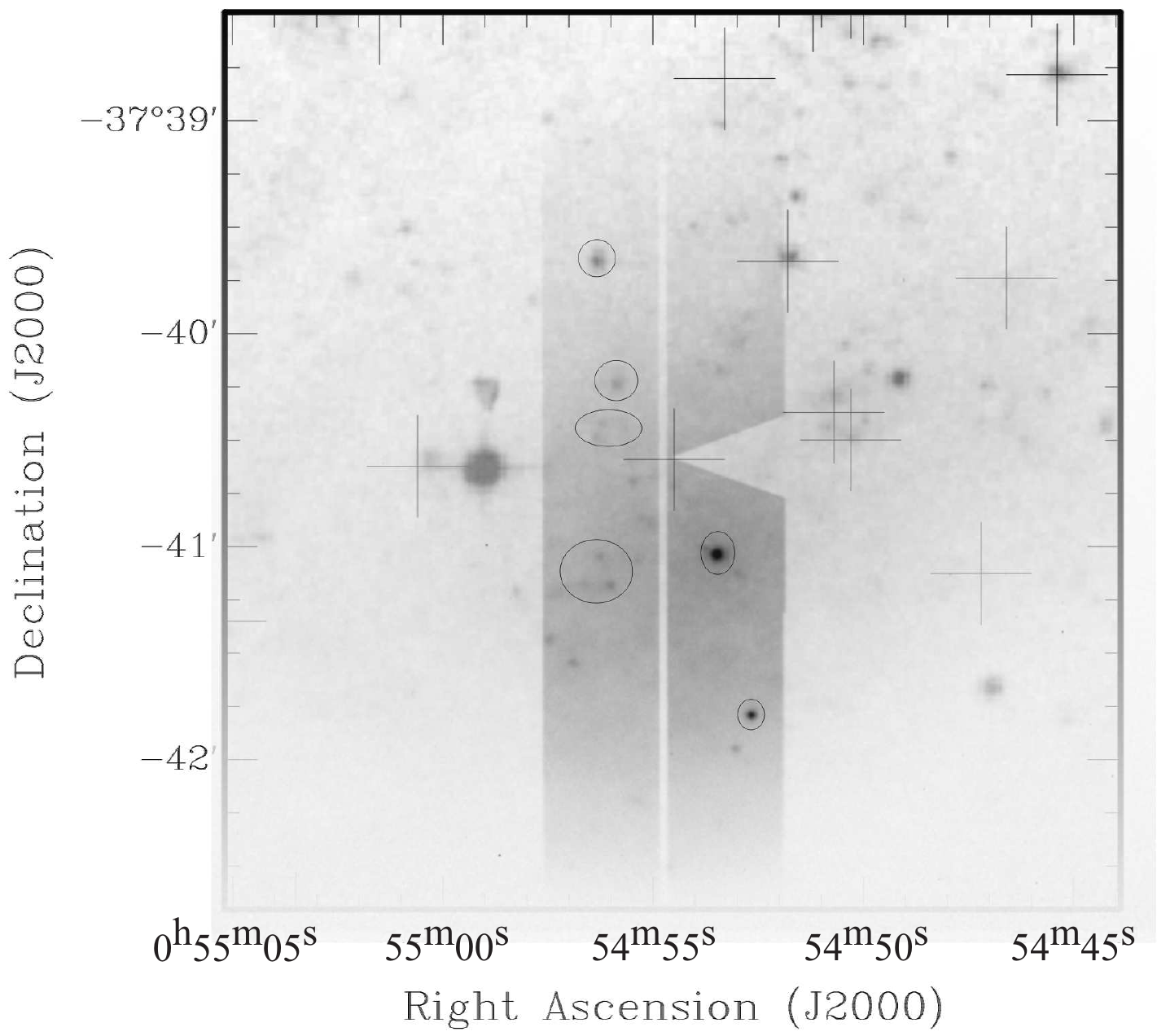}
{The DBS slit camera image laid on top of the MWF11 finding map for BL97 candidate \NS16. The circled objects are those which were used for the alignment of the two images. Due to pointing error the slit was not well aligned with the reported position of \NS16 but a \linrat\ of 0.70 (BL97) or \RatwErr{0.94}{0.06} (MWF11) was measured.}

\newpage

\figure[width=6in]{18}{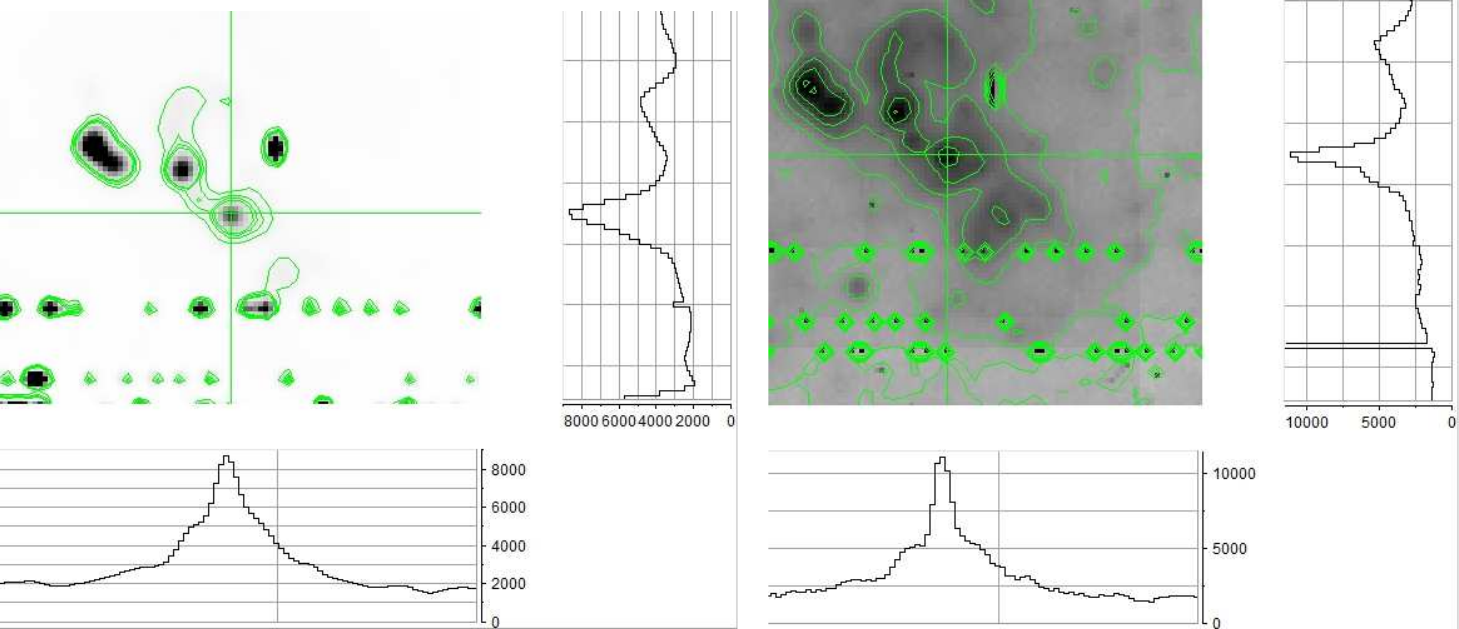}
{Stacked BL97 \Ha\ images of the SNR candidates. The left and right renditions differ in intensity scale, zoom level and contour line count. The profiles of the image stack for $x$ (E-W) and $y$ (N-S) axes are shown below and to the right, respectively. The lines across the rendition (crosshairs) indicate which CCD row ($x$) and column ($y$) are displayed in the profiles. These renditions were created with \emph{DS9} after stacking the SNR images with SBIG's \emph{CCDOps} software.}

\vspace{1cm}

\figure[width=6in]{19}{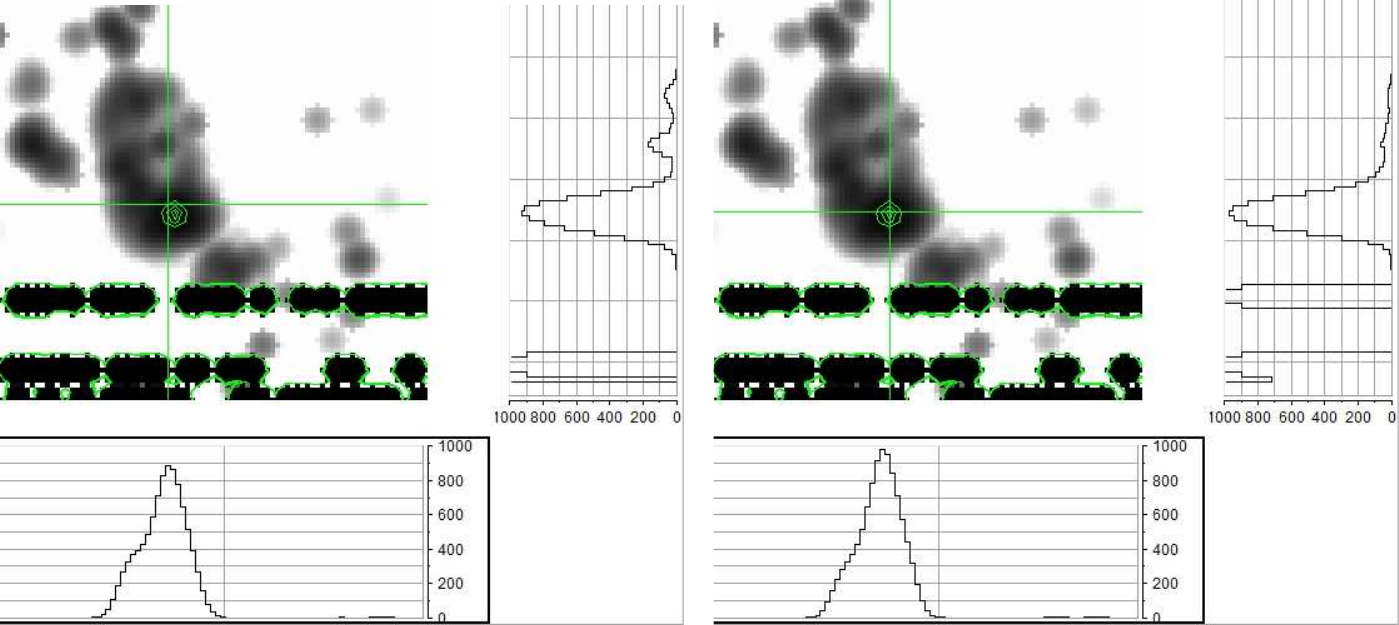}
{Stacked \Sii\ images of the BL97 SNR candidates. The left and right renditions differ in the takeoff location of the profiles. The left rendition's profiles are based on the position of the SNR candidate. The right rendition's profiles are based on the pixel with the maximum count. In each rendition, the profiles of the image stack for $x$ (E-W) and $y$ (N-S) axes are shown below and to the right, respectively. The lines (crosshairs) across the rendition indicate which CCD row ($x$) and column ($y$) are displayed in the profiles. These rendition were created with \emph{DS9} after stacking the SNR images with SBIG's \emph{CCDOps} software.}

\newpage

\figure{20}{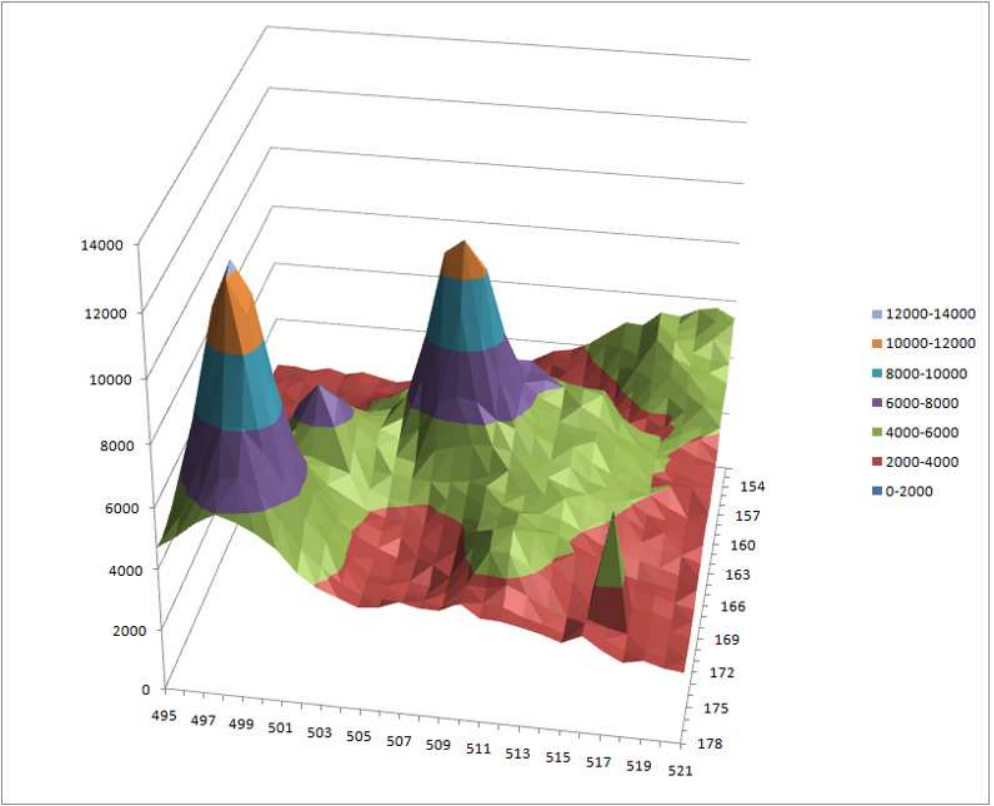}
{This is a $27\times27$ pixel ($110\times110$~pc) surface plot the central SNR location of the stacked \Ha\ images shown in Fig.~18. The data values of the central $9\times9$ region is shown in Table~10. A plot of the peak along the $x$-axis (E-W) at $y$-axis (N-S) pixel row 165 is shown in Fig.~21.}

\vspace{1cm}

\figure{21}{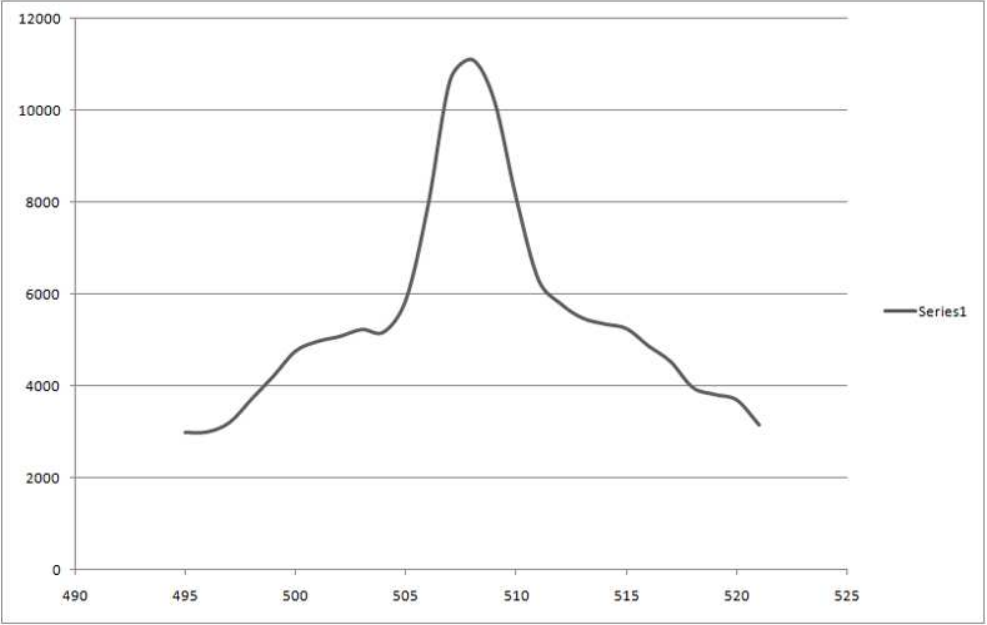}
{This is a plot of the stacked \Ha\ images $x$-axis (E-W) values in $y$-axis (N-S) row 165 shown in Fig.~20 and Table~10.}

\newpage

\figure{22}{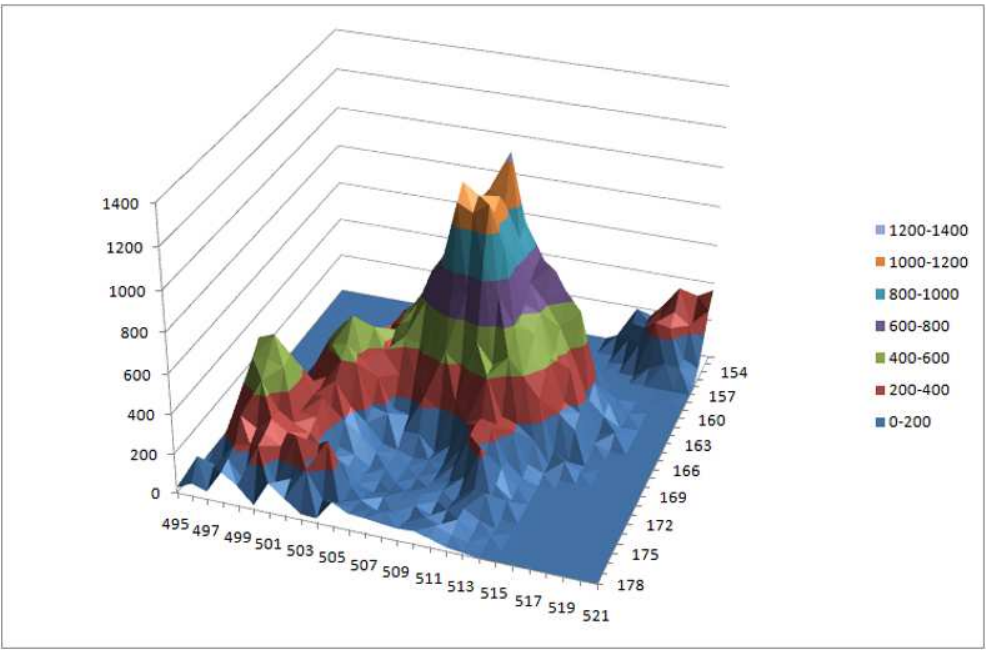}
{This is a $27\times27$ pixel ($110\times110$~pc) surface plot the central SNR location of the stacked \Sii\ images shown in Fig.~19. A plot of the peak along the $x$-axis (E-W) at $y$-axis (N-S) pixel row 165 is shown in Fig.~23.}

\vspace{1cm}

\figure{23}{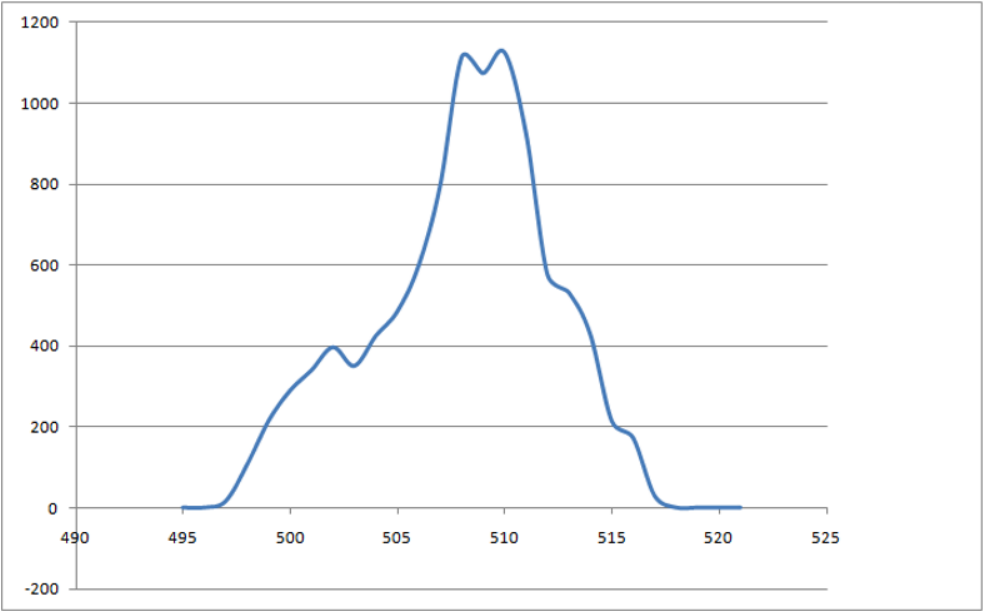}
{This is a plot of the stacked \Sii\ images $x$-axis values in $y$-axis row 165 shown in Fig.~22. The FWHM is taken at 550 counts as 7 pixels (28~pc).}

\newpage

\figure{24}{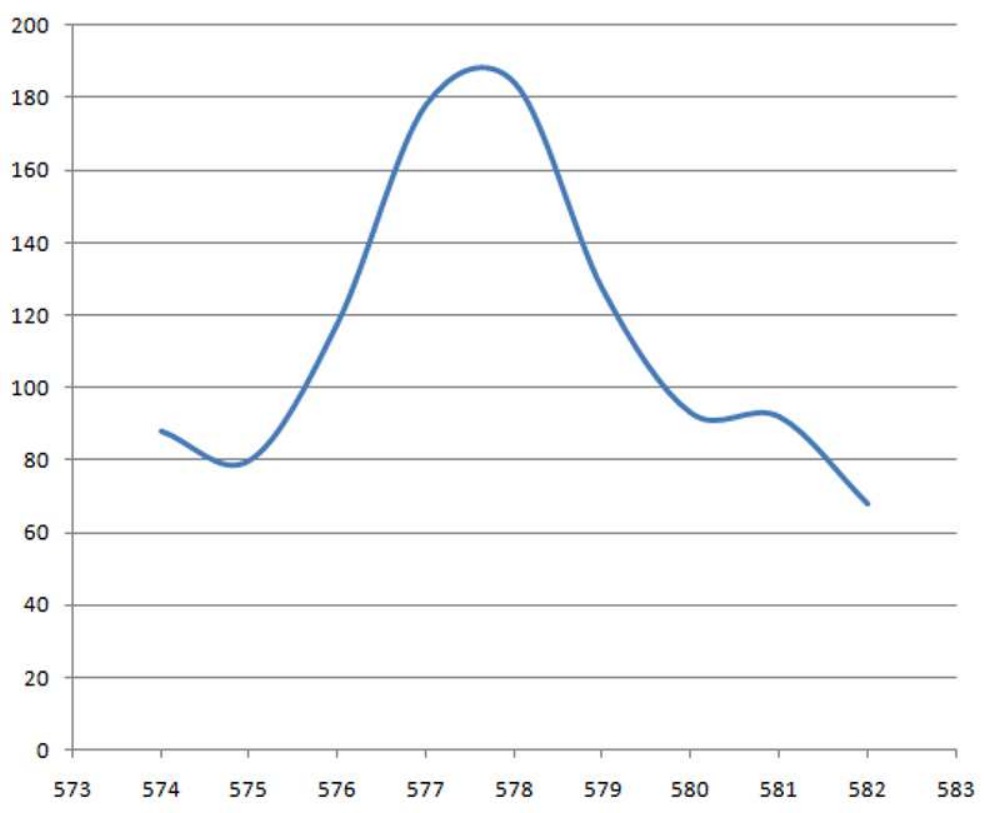}
{This is the $x$-axis (E-W) emission profile of a small star from the \Sii\ J image. The FWHM is taken at 130 count with a pixel width of 3.7 (15~pc).}

\vspace{1cm}

\figure[width=6.25in]{25}{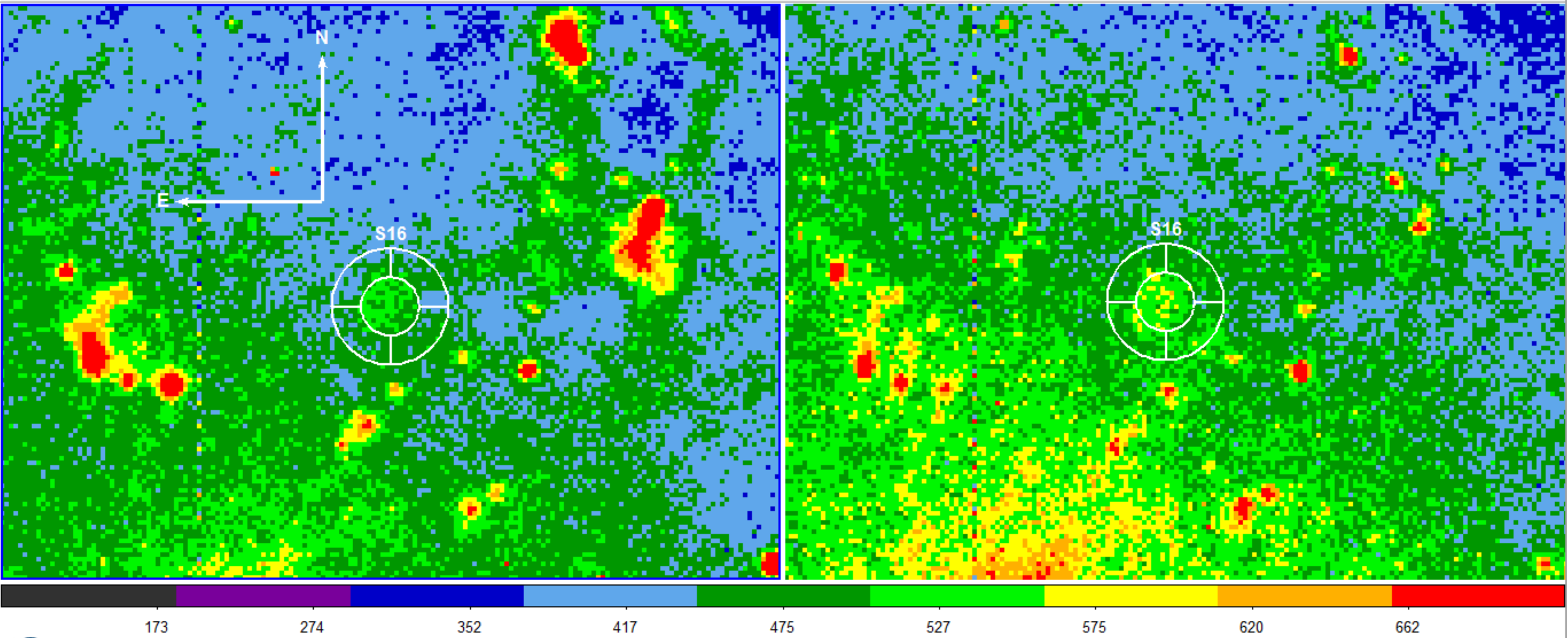}
{This is the BL97 location of \NS16 ($\alpha$ 00 54 54.46, $\delta$ $-37$ 40 35.46) as found on the BL97 images (G) with 2MASS calibrated astrometry. The \Ha\ image is on the left and the \Sii\ image is on the right. The inner circle of the panda is \Sec{5.1} (52~pc) in diameter. The color scale is AIPS0, Zoom=4, Scale = ZScale, Squared.}

\newpage

\figure{26}{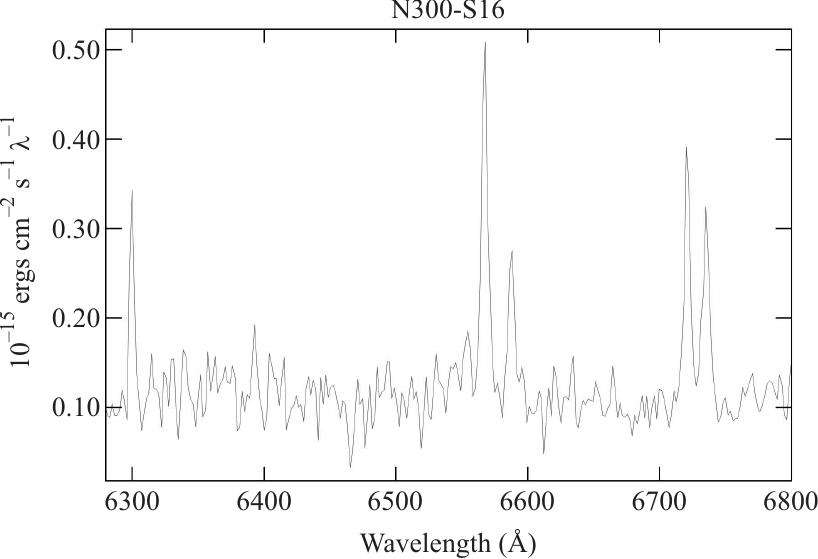}
{The spectrum of \NS16 from MWF11. The spectrum also contains a strong \Oi\ line at 6300~\AA\ which is an indicator of shock fronts typical of SNRs. Note the broad band noise level in the spectrum and the low flux density.}

\newpage

\figure{27}{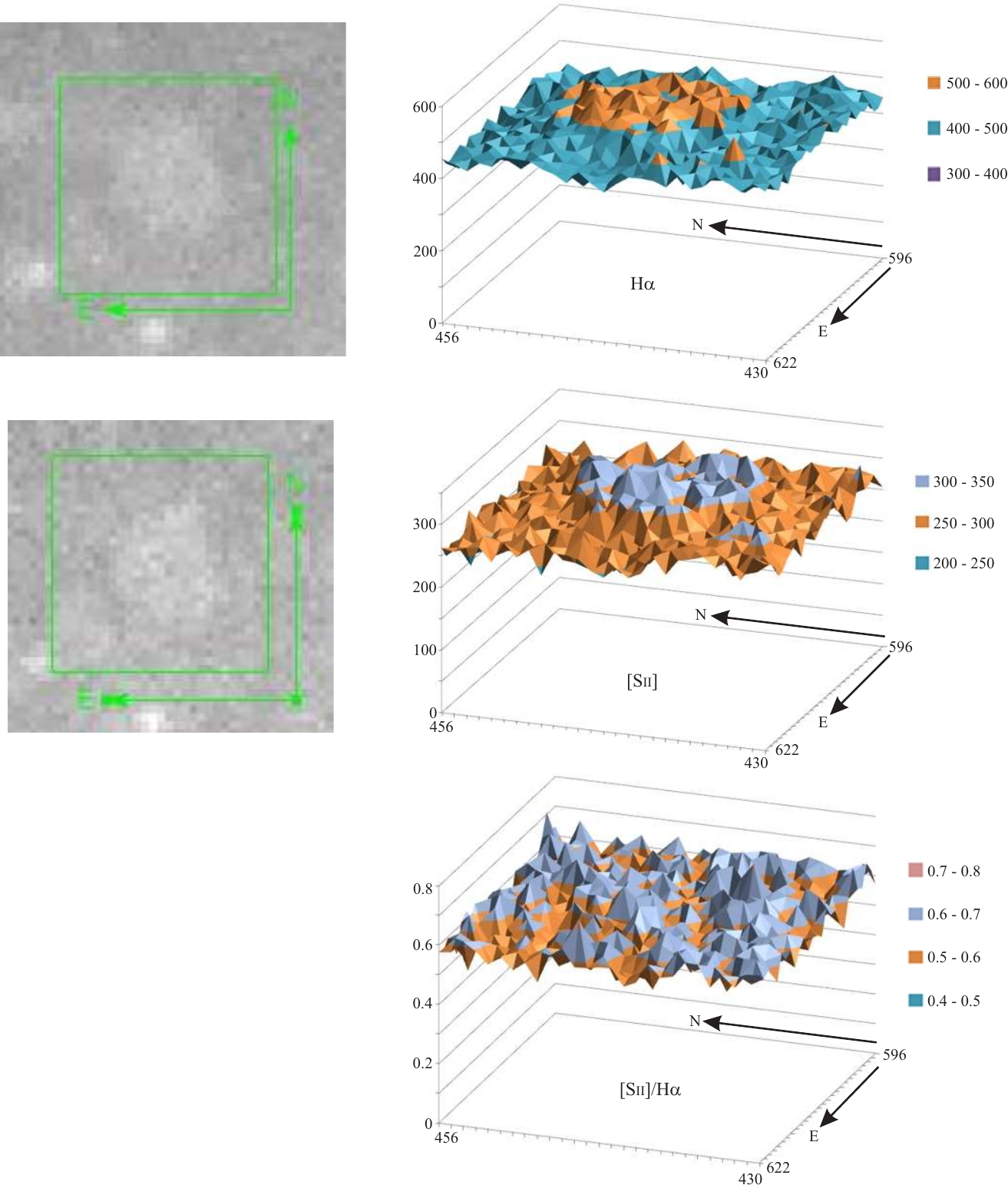}
{These are 3-D plots of the BL97 image CCD pixel counts of \NS16 in \Ha\ (top) and \Sii\ (middle). The pixels plotted (a 27 $\times$ 27 array centered on the candidate's coordinates) are shown by the green box in the image tile on the left. The plot is shown on the right. In both cases the emissions are only slightly above the background but clear discernible. The size of the candidate implies it to be at the end of its PDS stage. The 3-D plot on the bottom was the result of dividing the \Sii\ pixel value by the \Ha\ pixel value. The apparent SNR disappears in the high-level noise.}

%% file: Tables.tex

\newcommand\tablecaption[2]{\begin{centering}%
{\textbf{Table #1}. #2}\par%
\end{centering}}
\newcommand\tablehead[1]{\hline\hline #1 \\ \hline}
\newcommand\tablefoot{\hline\hline}
\newcommand\colhead[1]{\textbf{#1}}

\newpage


\tablecaption{1}{Gross Properties of \TG\ (from MWF11).}
\centerline{
\begin{tabular}{lll}
  \tablehead{\colhead{Property} & \colhead{Value} & \colhead{Reference}}
  Hubble Type                & SA(s)d                                & \citet{Tully}{1988} \\
                             &                                       & \citet{deVaucouleurs}{1991} \\
  R.A. (J2000.0)             & \RA{00}{54}{53.48}                    & NED \\
  Dec. (J2000.0)             & \DEC{-37}{41}{03.8}                   & NED \\
  Galactic Latitude          & \Deg{-77.17}                          & NED \\
  Radial Velocity            & 144 km/s (Solar)                      & \citet{Puche}{1990} \\
                             &                                       & \citet{Karachentsev}{2003} \\
  Inclination                & \Deg{46}                              & \citet{Tully}{1988} \\
                             & \Deg{42.6}                            & \citet{Puche}{1990} \\
  Distance                   & 2.1 Mpc                               & \citet{Freedman}{1992} \\
                             & 2.02 Mpc                              & \citet{Freedman}{2001} \\
                             & 1.88 Mpc                              & \citet{Bresolin}{2005}; \\
                             &                                       & \citet{Gieren}{2005} \\
  Observed Diameter ($D_{25}$) & 20.2 arcmin                         & \citet{Tully}{1988} \\
  Observed Diameter (UV isophotes) & 21.9 x 15.5 arcmin              & \citet{Gil de Paz}{2007} \\
  Galaxy Diameter            & 22.6 kpc, at 2.1 Mpc                  & Based on \citet{Gil de Paz}{2007} \\
  Mass (\Hi)                 & $2.4\times10^9\;$\msun                & \citet{Tully}{1988} \\
  $N_H$ Column Density       & $2.97\times10^{20}\;\mathrm{cm}^{-2}$ & \citet{Read}{1997} \\
  \tablefoot
  \multicolumn{3}{l}{Note. NED = NASA/IPAC Extragalactic Database (http://nedwww.ipac.caltech.edu/).}
\end{tabular}}

\newpage

\tablecaption{2}{Summary of observational results for the selected objects (from BL97, P04, and MWF11).\par The cut-in headers are from MWF11.}
\begin{centering}
\renewcommand\arraystretch{0.8}
\begin{tabular}{*8{>{\tiny}c}}
  \tablehead{
  \colhead{1}       & \colhead{2}      & \colhead{3} & \colhead{4} & \colhead{5}       & \colhead{6}       & \colhead{7} & \colhead{8}\\
  \colhead{Optical} & \colhead{Radio}  & RA          & Dec         & \colhead{\linrat} & \colhead{\linrat} & $\alpha\pm\Delta\alpha$ & \colhead{Diameter} \\
  \colhead{Object}  & \colhead{Object} & (h m s)     & (\deccolhead) & (BL97)            & (MWF11)           & (P04)    & \colhead{(pc, MWF11)}
  }
  \multicolumn{8}{c}{\tiny SNRs (MWF11)} \\
  \hline
  \NS1  &                   & 00 54 19.21 & $-37$ 37 23.96 & 0.44 & \RatwErr{0.46}{0.29} &  & 38 \\
  \NS2  &                   & 00 54 21.85 & $-37$ 40 27.11 & 0.49 & \RatwErr{0.72}{0.39} &  & 69 \\
  \NS4  &                   & 00 54 30.62 & $-37$ 40 53.75 & 0.71 & \RatwErr{0.93}{0.04} &  & 150 \\
  \NS5  &                   & 00 54 30.99 & $-37$ 37 33.96 & 0.46 & \RatwErr{0.56}{0.47} &  &  \\
  \NS6  & \ATCA{5431}{3825} & 00 54 31.91 & $-37$ 38 25.68 & 0.60 & 0.69                 & --$^a$ & 44 \\
  \NS7  &                   & 00 54 33.17 & $-37$ 40 16.90 & 0.57 & \RatwErr{0.57}{0.41} &  & 31 \\
  \NS8  &                   & 00 54 38.17 & $-37$ 41 14.88 & 0.61 & \RatwErr{0.58}{0.06} &  & 49 \\
  \NS9  &                   & 00 54 40.20 & $-37$ 41 02.12 & 0.44 & \RatwErr{0.53}{0.23} &  & 83 \\
  \NS12 &                   & 00 54 43.86 & $-37$ 43 39.08 & 0.52 & \RatwErr{0.73}{0.27} &  & 22 \\
  \NS13 &                   & 00 54 46.60 & $-37$ 39 44.32 & 0.59 & \RatwErr{0.91}{0.05} &  & 35 \\
  \NS14 &                   & 00 54 47.15 & $-37$ 41 07.63 & 0.63 & \RatwErr{1.08}{0.24} &  & 41 \\
  \NS15 &                   & 00 54 53.32 & $-37$ 38 48.24 & 0.65 & \RatwErr{0.57}{0.39} &  & 12 \\
  \NS16 &                   & 00 54 54.46 & $-37$ 40 35.46 & 0.70 & \RatwErr{0.94}{0.06} &  & 57 \\
  \NS17 &                   & 00 54 56.68 & $-37$ 43 57.70 & 0.69 & \RatwErr{0.96}{0.15} &  & 65 \\
  \NS19 &                   & 00 55 05.41 & $-37$ 41 21.04 & 0.53 & \RatwErr{0.70}{0.42} &  & 30 \\
  \NS20 &                   & 00 55 05.68 & $-37$ 46 13.35 & 0.75 & \RatwErr{0.79}{0.11} &  & 48 \\
  \NS22 &                   & 00 55 07.50 & $-37$ 40 43.20 & 0.27 & \RatwErr{0.46}{0.38} &  & 75 \\
  \NS24 &                   & 00 55 09.48 & $-37$ 40 46.21 & 0.80 & \RatwErr{0.64}{0.13} &  & 100 \\
  \NS25 &                   & 00 55 10.68 & $-37$ 41 27.13 & 0.64 & \RatwErr{0.54}{0.40} &  & 80 \\
  \NS26 & \ATCA{5515}{4439} & 00 55 15.46 & $-37$ 44 39.11 & 0.57 & \RatwErr{0.86}{0.67} & -- & 31 \\
  \NS27 &                   & 00 55 17.54 & $-37$ 44 36.65 & 0.70 & \RatwErr{0.64}{0.48} &  & 66 \\
  \NS28 & \ATCA{5533}{4314} & 00 55 33.76 & $-37$ 43 13.13 & 0.61 & \RatwErr{0.45}{0.15} & -- & 63 \\

  \hline
  \multicolumn{8}{c}{\tiny Other Objects (\Hii\ Regions? MWF11)} \\
  \hline

        & \ATCA{5438}{4144} & 00 54 38.16 & $-37$ 41 44.2  &      & \RatwErr{0.17}{0.07} & \RatwErr{-0.8}{0.2} & \\
        & \ATCA{5438}{4240} & 00 54 38.49 & $-37$ 42 40.5  &      & \RatwErr{0.25}{0.02} & -- & \\
        & \ATCA{5439}{3543} & 00 54 39.61 & $-37$ 35 43.4  &      & 0.27                 & \RatwErr{-0.4}{0.1} & \\
        & \ATCA{5441}{3348} & 00 54 41.05 & $-37$ 33 48.9  &      & 0.36                 & -- & \\
        & \ATCA{5442}{4313} & 00 54 42.70 & $-37$ 43 13.3  &      & \RatwErr{0.19}{0.07} & \RatwErr{-0.9}{0.3} & \\
        & \ATCA{5443}{4311} & 00 54 43.11 & $-37$ 43 11.0  &      & \RatwErr{0.18}{0.09} & \RatwErr{-0.6}{0.2} & \\
        & \ATCA{5445}{3847} & 00 54 45.39 & $-37$ 38 47.1  &      & \RatwErr{0.11}{0.03} & \RatwErr{-0.3}{0.1} & \\
        & \ATCA{5450}{4030} & 00 54 50.28 & $-37$ 40 30.0  &      & \RatwErr{0.32}{0.12} & \RatwErr{-0.5}{0.2} & \\
        & \ATCA{5450}{3822} & 00 54 50.30 & $-37$ 38 22.4  &      & \RatwErr{0.22}{0.14} & \RatwErr{-0.2}{0.2} & \\
        & \ATCA{5450}{4022} & 00 54 50.73 & $-37$ 40 22.2  &      & \RatwErr{0.38}{0.31} & \RatwErr{-0.3}{0.1} & 130 \\
        & \ATCA{5451}{3826} & 00 54 51.16 & $-37$ 38 26.1  &      & \RatwErr{0.26}{0.16} & \RatwErr{-1.2}{0.7} & \\
        & \ATCA{5451}{3939} & 00 54 51.79 & $-37$ 39 39.6  &      & \RatwErr{0.10}{0.18} & \RatwErr{-0.1}{0.2} & \\
        & \ATCA{5500}{4037} & 00 55 00.58 & $-37$ 40 37.4  &      & \RatwErr{0.25}{0.09} & \RatwErr{-0.4}{0.4} & \\
        & \ATCA{5501}{3829} & 00 55 01.49 & $-37$ 38 29.9  &      & \RatwErr{0.35}{0.12} & \RatwErr{-0.9}{0.1} & 31 \\
        & \ATCA{5503}{4246} & 00 55 03.50 & $-37$ 42 46.0  &      & \RatwErr{0.14}{0.04} & \RatwErr{-0.4}{0.1} & \\
        & \ATCA{5503}{4320} & 00 55 03.66 & $-37$ 43 20.1  &      & \RatwErr{0.15}{0.08} & \RatwErr{-0.7}{0.3} & \\
        & \ATCA{5512}{4140} & 00 55 12.70 & $-37$ 41 40.3  &      & \RatwErr{0.08}{0.02} & \RatwErr{-0.7}{0.1} & \\
  \NS3  &                   & 00 54 28.86 & $-37$ 41 53.32 & 0.40 & \RatwErr{0.24}{0.31} &  & 26 \\
  \NS10 & \ATCA{5440}{4049} & 00 54 40.87 & $-37$ 40 48.73 & 0.67 & \RatwErr{0.35}{0.15} & \RatwErr{-0.5}{0.3} & 63 \\
  \NS11 &                   & 00 54 42.54 & $-37$ 43 14.16 & 0.66 & \RatwErr{0.30}{0.12} &  & 150 \\
  \NS18 &                   & 00 55 01.39 & $-37$ 39 18.17 & 0.53 & \RatwErr{0.32}{0.32} &  & 69 \\
  \NS21 &                   & 00 55 07.15 & $-37$ 39 15.17 & 0.59 & \RatwErr{0.37}{0.30} &  & 41 \\
  \NS23 &                   & 00 55 09.10 & $-37$ 39 32.61 & 0.64 & \RatwErr{0.31}{0.08} &  & 43 \\

  \hline
  \multicolumn{8}{c}{\tiny No Signal (MWF11)} \\
  \hline

        & \ATCA{5423}{3648} & 00 54 23.84 & $-37$ 36 48.4 &  &  & \RatwErr{-0.7}{0.1} &  \\
        & \ATCA{5521}{4609} & 00 55 21.35 & $-37$ 46 09.6 &  &  & \RatwErr{-1.0}{0.3} &  \\
        & \ATCA{5523}{4632} & 00 55 23.95 & $-37$ 46 32.4 &  &  & \RatwErr{-0.9}{0.1} &  \\
        & \ATCA{5525}{3653} & 00 55 25.82 & $-37$ 36 53.8 &  &  & \RatwErr{-1.0}{0.1} &  \\
        & \ATCA{5528}{4903} & 00 55 28.25 & $-37$ 49 03.3 &  &  & \RatwErr{-0.6}{0.3} &  \\
        & \ATCA{5541}{4033} & 00 55 41.94 & $-37$ 40 33.5 &  &  & -- &  \\

  \tablefoot
  \multicolumn{8}{l}{\tiny $^a$Spectral index not determined.}
\end{tabular}
\end{centering}

\newpage

\tablecaption{3}{Archival HST files organized by filter center wavelength.}
\begin{centering}
\renewcommand\arraystretch{0.7}
\begin{tabular}{*4{>{\tiny}c}}
  \tablehead{
  \colhead{1}                & \colhead{2}     & \colhead{3}           & \colhead{4} \\
  \colhead{Wavelength (\AA)} & \colhead{Files} & \colhead{Proposal ID} & \colhead{Associated Publications}
  }
  2993 & \texttt{u8hhvp02m\_drz} & 9677 & \citet{Wadadekar}{2006}. \\

  4318 & \texttt{j8d702010\_drz} & 9492 & \citet{Gliozzi}{2009}; \citet{Gogarten}{2009b}; \\
       & \texttt{j8d702mcq\_flt} &      & \citet{Nantais}{2010}; \citet{Gogarten}{2010}; \\
       & \texttt{j8d702mjq\_flt} &      & \citet{Gieren}{2004}; \citet{Bresolin}{2005}; \\
       & \texttt{j8d702mrq\_flt} &      & \citet{Rizzi}{2007}; \citet{Kudritzki}{2008}; \\
       & \texttt{j8d703010\_drz} &      & \citet{Bond}{2009}; \citet{Barth}{2009}; \\
       & \texttt{j8d703viq\_flt} &      & \citet{Dalcanton}{2009}; \citet{Rizzi}{2006}; \\
       & \texttt{j8d703vpq\_flt} &      & \citet{Tikhonov}{2005}; \\
       & \texttt{j8d703vxq\_flt} &      & \citet{Kuntz}{2010}. \\

  4747 & \texttt{j9ra11010\_drz} & 10915 & \citet{Berger}{2009}; \citet{Gogarten}{2009a}; \\
       & \texttt{j9ra11neq\_flt} &       & \citet{Holwerda}{2009}; \citet{Lianou}{2009}; \\
       & \texttt{j9ra11nfq\_flt} &       & \citet{Nantais}{2010}; \citet{Williams}{2010}; \\
       &                         &       & \citet{Melbourne}{2010}; \citet{Gogarten}{2010}; \\
       &                         &       & \citet{de Mello}{2008}; \citet{Girardi}{2008}; \\
       &                         &       & \citet{Mould}{2008}; \citet{Bond}{2009}; \\
       &                         &       & \citet{Williams}{2009}; \citet{Kornei}{2009}; \\
       &                         &       & \citet{Williams}{2009}. \\

  5360 & \texttt{j8d702020\_drz} & 9492  & See \wleq{4318}. \\
       & \texttt{j8d702mdq\_flt} \\
       & \texttt{j8d702mlq\_flt} \\
       & \texttt{j8d702mtq\_flt} \\
       & \texttt{j8d703020\_drz} \\
       & \texttt{j8d703vjq\_flt} \\
       & \texttt{j8d703vrq\_flt} \\
       & \texttt{j8d703vzq\_flt} \\

  5484 & \texttt{u6713701m\_drz} & 8591 & \citet{Graham}{2001}; \citet{Schinnerer}{2003}; \\
       & \texttt{u6713702r\_drz} &      & \citet[]{Larsen}{2004}; \citet{Lauer}{2005}; \\
       & \texttt{u6713703r\_drz} &      & \citet{Rosolowsky}{2005}; \citet{Lauer}{2007a}; \\
       & \texttt{u6713704r\_drz} &      & \citet{Lauer}{2007b}; \citet{Gonz{\'a}lez Delgado}{2008}; \\
       &                         &      & \citet{Dai}{2008}; \citet{Siopis}{2009}; \\
       &                         &      & \citet{Beifiori}{2009}. \\

  5741 & \texttt{o6j3bzkyq\_flt} & 9285 & \citet{Shaw}{2006}. \\

  5921 & \texttt{j9ra11020\_drz} & 10915 & See \wleq{4747}. \\
       & \texttt{j9ra11nhq\_flt} \\
       & \texttt{j9ra11njq\_flt} \\

  6001 & \texttt{u8hhvp01m\_drz} & 9677 & \citet{Wadadekar}{2006}. \\

  6001 & \texttt{u8ixhd01m\_drz} & 9676 & \citet{Tikhonov}{2009}; \citet{Van Dyk}{2003}; \\
       & \texttt{u8ixhd02m\_drz} &      & \citet[]{Larsen}{2004}; \citet{Smartt}{2004}; \\
       &                         &      & \citet{Tikhonov}{2005a}; \citet[]{Sugerman}{2005}; \\
       &                         &      & \citet{Milone}{2006}; \citet{Wadadekar}{2006}; \\
       &                         &      & \citet{Sugerman}{2006}; \citet{Shaw}{2007}; \\
       &                         &      & \citet{Maund}{2009}; \citet{Guerrero}{2008}. \\

  6564 (\Ha) & \texttt{u6713705r\_drz} & 8591  & See \wleq{5484}. \\
       & \texttt{u6713706r\_drz} \\
       & \texttt{u6713707r\_drz} \\
       & \texttt{u6713708r\_drz} \\
       & \texttt{u6713709r\_drz} \\

  7996 & \texttt{u65w0201r\_drz} & 8599  & \citet{Windhorst}{2002}; \citet{B{\"o}ker}{2003a}; \\
       & \texttt{u65w0202r\_drz} &       & \citet[]{Larsen}{2004}; \citet{Butler}{2004} \\
       & \texttt{u65w0203r\_drz} &       & \citet{Walcher}{2005}; \citet{de Grijs}{2005}; \\
       &                         &       & \citet{Tully}{2006}; \citet{Rossa}{2006}; \\
       &                         &       & \citet{Schinnerer}{2006}; \citet{Peeples}{2006}; \\
       &                         &       & \citet{Ganda}{2006}; \citet{Cao}{2007}; \\
       &                         &       & \citet{Seth}{2008}; \citet{B{\"o}ker}{2003b}; \\
       &                         &       & \citet{Gonz{\'a}lez Delgado}{2008}; \citet{Ghosh}{2009}; \\
       &                         &       & \citet{B{\"o}ker}{2004}; \citet{Andersen}{2008}; \\
       &                         &       & \citet{B{\"o}ker}{2002}. \\
  8057 & \texttt{j9ra11030\_drz} & 10915 & See \wleq{4747}. \\
       & \texttt{j9ra11nlq\_flt} \\
       & \texttt{j9ra11nnq\_flt} \\

  8060 & \texttt{j8d702030\_drz} & 9492  & See \wleq{4318}. \\
       & \texttt{j8d702mfq\_flt} \\
       & \texttt{j8d702moq\_flt} \\
       & \texttt{j8d702mvq\_flt} \\
       & \texttt{j8d702n0q\_drz} \\
       & \texttt{j8d702n0q\_flt} \\
       & \texttt{j8d703030\_drz} \\
       & \texttt{j8d703vlq\_flt} \\
       & \texttt{j8d703vuq\_flt} \\
       & \texttt{j8d703w1q\_flt} \\
       & \texttt{j8d703w6q\_drz} \\
       & \texttt{j8d703w6q\_flt} \\
  \tablefoot
\end{tabular}
\end{centering}

\newpage

\tablecaption{4}{Central Wavelengths of HST Image File Sequences.}
\centerline{
\renewcommand\arraystretch{0.75}
\begin{tabular}{*3{>{\tiny}l}}
  \tablehead{
  \colhead{1} & \colhead{2} & \colhead{3} \\
                             &                               & \colhead{Known SNR Ions/Atoms with Wavelengths} \\
  \colhead{Wavelength (\AA)} & \colhead{RMS Bandwidth (\AA)} & \colhead{\AA, Fesen et al. 1996}
  }
  2993 & 325  & \ion{Mg}{i}: 2852.13 \\
  4318 & 293  & \Hi: 4340.49 (H$\gamma$)\\
       &      & \ion{C}{ii}: 4267.00, 4267.26 \\
       &      & \Oiii: 4363.21 \\
       &      & \fion{Fe}{ii}: 4177.21, 4243.98, \\
       &      & 4244.81, 4287.40, 4358.10, 4358.37, 4359.34, \\
       &      & 4413.78, 4414.45, 4416.27, 4452.11, 4457.95 \\
       &      & \fion{Fe}{v}: 4227.20 \\
  4747 & 420  & \ion{H}{i}: 4861.36 (H$\beta$) \\
       &      & \ion{He}{i}: 4713.14, 4713.37, 4921.93 \\
       &      & \ion{He}{ii}: 4685.68 \\
       &      & \fion{Ne}{iv}: 4724.17, 4725.60 \\
       &      & \ion{Mg}{i}: 4571.10 \\
       &      & \fion{Mg}{i}: 4562.48 \\
       &      & \fion{Ar}{iv}: 4711.33, 4740.20 \\
       &      & \fion{Fe}{ii}: 4632.27, 4774.74, 4814.55, \\
       &      & 4889.63, 4889.70, 4905.35 \\
       &      & \fion{Fe}{iii}: 4658.10, 4701.62, 4733.93, \\
       &      & 4754.83, 4769.60, 4777.88, 4813.90, 4881.11, \\
       &      & 4924.50, 4930.50 \\
       &      & \fion{Fe}{vii}: 4893.40, 4942.50 \\
  5360 & 360  & \ion{He}{ii}: 5411.52 \\
       &      & \fion{N}{i}: 5197.90, 5200.26 \\
       &      & \fion{Cl}{iii}: 5517.71 \\
       &      & \fion{Ar}{iii}: 5191.82 \\
       &      & \fion{Ca}{v}: 5309.18 \\
       &      & \fion{Fe}{ii}: 5184.80, 5220.06, 5261.61, \\
       &      & 5268.88, 5273.38, 5296.84, 5333.65, 5376.47, \\
       &      & 5412.64, 5413.34, 5527.33 \\
       &      & \fion{Fe}{iii}: 5270.30 \\
       &      & \fion{Fe}{vi}: 5277.80, 5335.20, 5424.20, 5426.60, 5484.80 \\
       &      & \fion{Fe}{xiv}: 5302.86 \\
  5484 & 206  & \ion{He}{ii}: 5411.52 \\
       &      & \fion{O}{i}: 5577.34 \\
       &      & \fion{Cl}{iii}: 5517.71 \\
       &      & \fion{Fe}{ii}: 5412.64, 5413.34, 5527.33 \\
       &      & \fion{Fe}{vi}: 5424.20, 5426.60, 5484.80 \\
  5741 & 1836 & \ion{H}{i}: 4861.36 (H$\beta$), 6562.85 (\Ha) \\
       &      & \ion{He}{i}: 4921.93, 5015.68, (5876) \\
       &      & \ion{He}{ii}: 5411.52 \\
       &      & \fion{N}{i}: 5197.90, 5200.26 \\
       &      & \fion{N}{ii}: 5754.59, 6548.05, 6583.45 \\
       &      & \fion{O}{i}: 5577.34, 6300.30, 6363.78 \\
       &      & \fion{O}{iii}: 4958.91, 5006.84 \\
       &      & \ion{Na}{i}: 5889.95, 5895.92 \\
       &      & \fion{S}{iii}: 6312.06 \\
       &      & \fion{Cl}{iii}: 5517.71 \\
       &      & \fion{Ar}{iii}: 5191.82 \\
       &      & \fion{Ar}{v}: 6435.10 \\
       &      & \fion{Ca}{v}: 5309.18 \\
       &      & \fion{Fe}{ii}: 4889.63, 4889.70, 4905.35, \\
       &      & 4973.39, 5039.10, 5043.53, 5072.40, 5107.95, \\
       &      & 5111.63, 5158.00, 5158.81, 5184.80, 5220.06, \\
       &      & 5261.61, 5268.88, 5273.38, 5296.84, 5333.65, \\
       &      & 5376.47, 5412.64, 5413.34, 5527.33 \\
       &      & \fion{Fe}{iii}: 4881.11, 4924.50, 4930.50, \\
       &      & 4985.90, 4987.20, 5270.30 \\
       &      & \fion{Fe}{vi}: 4972.50, 5145.80, 5176.00, \\
       &      & 5277.80, 5335.20, 5424.20, 5426.60, 5484.80, \\
       &      & 5631.10, 5677.00 \\
       &      & \fion{Fe}{vii}: 4893.40, 4942.50, 5158.90, 5720.70, 6087.00 \\
       &      & \fion{Fe}{x}: 6374.51 \\
       &      & \fion{Fe}{xiv}: 5302.86 \\
  \hline
  \multicolumn{3}{l}{\tiny$^a$Unable to find these wavelengths for \Hi\ in NIST database.} \\
  \multicolumn{3}{r}{\tiny Continued}
\end{tabular}}

\newpage

\tablecaption{4}{Central Wavelengths of HST Image File Sequences (continued).}
\centerline{
\renewcommand\arraystretch{0.75}
\begin{tabular}{*3{>{\tiny}l}}
  \tablehead{
  \colhead{1} & \colhead{2} & \colhead{3} \\
                             &                               & \colhead{Known SNR Ions/Atoms with Wavelengths} \\
  \colhead{Wavelength (\AA)} & \colhead{RMS Bandwidth (\AA)} & \colhead{\AA, Fesen et al. 1996}
  }
  5921 & 672  & \ion{He}{i}: (5876) \\
       &      & \fion{N}{ii}: 5754.59 \\
       &      & \ion{Na}{i}: 5889.95, 5895.92 \\
       &      & \fion{Fe}{vi}: 5631.10, 5677.00 \\
       &      & \fion{Fe}{vii}: 5720.70, 6087.00 \\
  6001 & 638  & \ion{He}{i}: (5876) \\
       &      & \fion{N}{ii}: 5754.59 \\
       &      & \fion{O}{i}: 6300.30 \\
       &      & \ion{Na}{i}: 5889.95, 5895.92 \\
       &      & \fion{S}{iii}: 6312.06 \\
       &      & \fion{Fe}{vii}: 5720.70, 6087.00 \\
  6564 & 54   & \ion{H}{i}: 6562.85 (\Ha) \\
       &      & \fion{N}{ii}: 6548.05, 6583.45 \\
  7996 & 646  & \ion{He}{ii}: 8236.77 \\
       &      & \ion{O}{i}: (7774) \\
       &      & \fion{Ar}{iii}: 7751.06 \\
       &      & \fion{Cr}{ii}: 7999.85, 8125.22, 8229.55, 8308.39 \\
       &      & \fion{Fe}{ii}: 7686.19, 7686.90 \\
       &      & \fion{Fe}{xi}: 7891.80 \\
       &      & \fion{Ni}{ii}: 8300.99 \\
  8057 & 652  & \ion{H}{i}: 8345.55, 8359.00, 8374.48$^a$ \\
       &      & \ion{He}{ii}: 8236.77 \\
       &      & \ion{O}{i}: (7774) \\
       &      & \fion{Ar}{iii}: 7751.06 \\
       &      & \fion{Cr}{ii}: 7999.85, 8125.22, 8229.55, 8308.39, 8357.51 \\
       &      & \fion{Fe}{xi}: 7891.80 \\
       &      & \fion{Ni}{ii}: 8300.99 \\
  8060 & 653  & \ion{H}{i}: 8345.55, 8359.00, 8374.48$^a$ \\
       &      & \ion{He}{ii}: 8236.77 \\
       &      & \ion{O}{i}: (7774) \\
       &      & \fion{Ar}{iii}: 7751.06 \\
       &      & \fion{Cr}{ii}: 7999.85, 8125.22, 8229.55, 8308.39, 8357.51 \\
       &      & \fion{Fe}{xi}: 7891.80 \\
       &      & \fion{Ni}{ii}: 8300.99 \\
  \tablefoot
  \multicolumn{3}{l}{\tiny$^a$Unable to find these wavelengths for \Hi\ in NIST database.}
\end{tabular}}

\newpage

\tablecaption{5}{HST filters containing ion/atom species important for SNR identification.}
\renewcommand\arraystretch{1.0}
\centerline{
\begin{tabular}{*3{>{\small}{c}}}
  \tablehead{
  \colhead 1 & \colhead 2 & \colhead 3 \\
  \colhead{Wavelength (\AA)} & \colhead{RMS Bandwidth (\AA)} & \colhead{SNR Diagnostic Ions/Atoms with Wavelengths (\AA)}
  }
  5741 & 1836 & \Hi: 4861.36 (\Hb), 6562.85 (\Ha) \\
       &      & \Nii: 5754.59, 6548.05, 6583.45 \\
       &      & \Oi: 5577.34, 6300.30, 6363.78 \\
       &      & \Oiii: 4958.91, 5006.84 \\
  6001 & 638  & \Oi: 6300.30 \\
  6564 & 54   & \Hi: 6562.85 (\Ha) \\
       &      & \Nii: 6548.05, 6583.45 \\
  \tablefoot
\end{tabular}}

\vspace{1cm}

\tablecaption{6}{Five candidates found in the \Ha\ HST image \texttt{u6713709r\_drz.fits}. The positions are the J2000 coordinates reported by P04 (radio sources) and BL97 (optical candidates).}
\centerline{
\renewcommand\arraystretch{0.8}
\begin{tabular}{*8{>{\tiny}c}}
  \tablehead{
  \colhead{1}       & \colhead{2}      & \colhead{3} & \colhead{4} & \colhead{5}       & \colhead{6}       & \colhead{7} & \colhead{8}\\
  \colhead{Optical} & \colhead{Radio}  & RA          & Dec         & \colhead{\linrat} & \colhead{\linrat} & $\alpha\pm\Delta\alpha$ & \colhead{Diameter} \\
  \colhead{Object}  & \colhead{Object} & (h m s)     & (\deccolhead) & (BL97)            & (MWF11)           & (P04)    & \colhead{(pc, MWF11)}
  }
        & \ATCA{5450}{4030} & 00 54 50.28 & $-37$ 40 30.0  &      & \RatwErr{0.32}{0.12} & \RatwErr{-0.5}{0.2} & \\
        & \ATCA{5450}{4022} & 00 54 50.73 & $-37$ 40 22.2  &      & \RatwErr{0.38}{0.31} & \RatwErr{-0.3}{0.1} & 130 \\
        & \ATCA{5451}{3939} & 00 54 51.79 & $-37$ 39 39.6  &      & \RatwErr{0.10}{0.18} & \RatwErr{-0.1}{0.2} & \\
        & \ATCA{5500}{4037} & 00 55 00.58 & $-37$ 40 37.4  &      & \RatwErr{0.25}{0.09} & \RatwErr{-0.4}{0.4} & \\
  \NS16 &                   & 00 54 54.46 & $-37$ 40 35.46 & 0.70 & \RatwErr{0.94}{0.06} &  & 57 \\
  \tablefoot
  \multicolumn{8}{l}{\tiny $^a$Spectral index not determined.}
\end{tabular}}

\vspace{1cm}

\tablecaption{7}{Image center from BL97 and 2MASS calibrated.}
\centerline{
\begin{tabular}{l*3{c}}
  \tablehead{
  \colhead1       & \colhead2                        & \colhead3                     & \colhead4 \\
                  & \colhead{BL97 Center}            & \colhead{Calibrated Center}   & \colhead{Difference} \\
  \colhead{Image} & \colhead{(h m s \deccolhead)}    & \colhead{(h m s \deccolhead)} & \colhead{(\Sec{})}
  }
  D              & 00 54 23.030 $-37$ 35 50.600      & 00 54 22.966 $-37$ 35 51.374  & 1.081 \\
  E              & 00 55 38.770 $-37$ 40 47.900      & 00 55 38.793 $-37$ 40 48.215  & 0.420 \\
  F              & 00 55 13.460 $-37$ 40 51.900      & 00 55 13.550 $-37$ 40 52.086  & 1.082 \\
  G              & 00 54 47.360 $-37$ 40 50.400      & 00 54 47.359 $-37$ 40 51.305  & 0.905 \\
  H              & 00 54 22.840 $-37$ 40 50.700      & 00 54 22.856 $-37$ 40 52.042  & 1.356 \\
  J              & 00 55 14.000 $-37$ 45 50.300      & 00 55 13.987 $-37$ 45 50.463  & 0.227 \\
  \tablefoot
\end{tabular}}

\newpage

\tablecaption{8}{Calculated pixel positions of BL97 candidate SNRs.}
\centerline{
\begin{tabular}{l*3{c}}
  \tablehead{
  \colhead1 & \colhead2                     & \colhead3   & \colhead4 \\
            & \colhead{BL97 Coordinates}    &       & \colhead{$(x,y)$ Location} \\
  \colhead{Object}   & \colhead{(h m s \deccolhead)} & \colhead{Image} & \colhead{(pixel position)}
  }
  N300-S1  & 00 54 19.21 $-37$ 37 23.96 & D     & 508, 165 \\
  N300-S2  & 00 54 21.85 $-37$ 40 27.11 & H     & 431, 458 \\
  N300-S3  & 00 54 28.86 $-37$ 41 51.32 & H     & 219, 254 \\
  N300-S4  & 00 54 30.62 $-37$ 40 53.75 & H     & 170, 398 \\
  N300-S5  & 00 54 30.99 $-37$ 37 33.96 & D     & 158, 149 \\
  N300-S6  & 00 54 31.91 $-37$ 38 25.68 & D     & 128, 020 \\
           &                            & H     & 141, 767 \\
  N300-S7  & 00 54 33.17 $-37$ 40 16.90 & H     & 097, 492 \\
  N300-S8  & 00 54 38.17 $-37$ 41 14.88 & G     & 671, 333 \\
  N300-S9  & 00 54 40.20 $-37$ 41 02.12 & G     & 612, 366 \\
  N300-S10 & 00 54 40.87 $-37$ 40 48.73 & G     & 593, 400 \\
  N300-S11 & 00 54 42.54 $-37$ 43 14.16 & G     & 535, 040 \\
  N300-S12 & 00 54 43.86 $-37$ 43 39.08 & (I) \\
  N300-S13 & 00 54 46.60 $-37$ 39 44.32 & G     & 427, 564 \\
  N300-S14 & 00 54 47.15 $-37$ 41 07.63 & G     & 406, 357 \\
  N300-S15 & 00 54 53.32 $-37$ 38 48.24 & G     & 231, 709 \\
  N300-S16 & 00 54 54.46 $-37$ 40 35.46 & G     & 191, 443 \\
  N300-S17 & 00 54 56.68 $-37$ 43 57.70 & (I) \\
  N300-S18 & 00 55 01.39 $-37$ 39 18.17 & F     & 762, 629 \\
  N300-S19 & 00 55 05.41 $-37$ 41 21.04 & F     & 636, 323 \\
  N300-S20 & 00 55 05.68 $-37$ 46 13.35 & J     & 645, 336 \\
  N300-S21 & 00 55 07.15 $-37$ 39 15.17 & F     & 592, 639 \\
  N300-S22 & 00 55 07.50 $-37$ 40 43.18 & F     & 577, 419 \\
  N300-S23 & 00 55 09.10 $-37$ 39 32.61 & F     & 535, 597 \\
  N300-S24 & 00 55 09.48 $-37$ 40 46.21 & F     & 518, 413 \\
  N300-S25 & 00 55 10.68 $-37$ 41 27.13 & F     & 482, 311 \\
  N300-S26 & 00 55 15.46 $-37$ 44 39.31 & J     & 361, 579 \\
  N300-S27 & 00 55 17.54 $-37$ 44 36.65 & J     & 299, 587 \\
  N300-S28 & 00 55 33.76 $-37$ 43 13.33 & E     & 541, 034 \\
  \tablefoot
\end{tabular}}

\newpage

\tablecaption{9}{A comparison of positions of a bright star in the image sets. Because the BL97 images were calibrated to 2MASS those star coordinates are used as the standard.}
\centerline{
\begin{tabular}{l*5{c}}
  \tablehead{
  \colhead 1 & \colhead 2 & \colhead 3 & \colhead 4 & \colhead 5 & \colhead 6 \\
  \colhead{Image Set} & \colhead{RA (J2000)} & \colhead{$\Delta$ (\Sec{})} & \colhead{Dec (J2000)} & \colhead{$\Delta$ (\Sec{})} & \colhead{Distance (\Sec{})}
  }
  BL97 (Image G)                          & 00 54 49.146 & ---      & $-37$ 40 11.99 & ---  & --- \\
  HST File (\texttt{u6713709r\_drz.fits}) & 00 54 49.085 & $-0.061$ & $-37$ 40 12.13 & 0.14 & 0.15 \\
  DSS2-Red                                & 00 54 49.099 & $-0.047$ & $-37$ 40 12.11 & 0.12 & 0.13 \\
  \tablefoot
\end{tabular}}

\vspace{1cm}

\tablecaption{10}{A $9\times9$ region of CCD pixel values ($37\times37$~pc) surrounding the central \Ha\ peak in Fig.~18. The $x$-axis (E-W) pixel numbers are across the top and the $y$-axis (N-S) pixel numbers are along the left side of the table. A surface plot of these data is shown in the central region of Fig.~20 and a line plot of the peak is shown in Fig.~21.}
\centerline{
\begin{tabular}{l|*9{r}}
  \tablehead{
    & \colhead{504} & \colhead{505} & \colhead{506} & \colhead{507} & \colhead{508} & \colhead{509} & \colhead{510} & \colhead{511} & \colhead{512}
  }
  \textbf{169} & 4664 & 4606 & 4489 &  4482 &  4507 &  4610 & 4842 & 5088 & 5071 \\
  \textbf{168} & 5168 & 4856 & 4704 &  5165 &  5357 &  5485 & 5495 & 5369 & 5272 \\
  \textbf{167} & 5400 & 5122 & 5478 &  6374 &  6677 &  6588 & 6046 & 5802 & 5665 \\
  \textbf{166} & 5270 & 5494 & 6869 &  8522 &  9109 &  8479 & 7062 & 6104 & 5669 \\
  \textbf{165} & 5182 & 5889 & 7939 & 10668 & 11117 & 10225 & 8087 & 6320 & 5801 \\
  \textbf{164} & 5088 & 5611 & 7712 &  9893 & 10624 &  9952 & 7889 & 6366 & 5851 \\
  \textbf{163} & 4897 & 5182 & 6255 &  7615 &  7955 &  7798 & 7071 & 6165 & 6185 \\
  \textbf{162} & 4775 & 4922 & 5444 &  5919 &  6285 &  6397 & 6471 & 6066 & 6040 \\
  \textbf{161} & 4709 & 4740 & 5362 &  5681 &  6046 &  6129 & 6120 & 6164 & 5652 \\
  \tablefoot
\end{tabular}}